\documentclass{aa}  

\usepackage{graphicx}
\usepackage{txfonts}
\usepackage{hyperref}
\usepackage{siunitx}
\usepackage{soul}
\usepackage[dvipsnames]{xcolor}
\usepackage{natbib}
\usepackage{float}

\def\ts     {\thinspace} 
\usepackage{amsmath}

\def\kms  {\ifmmode{{\rm \ts km\ts s}^{-1}}\else{\ts km\ts s$^{-1}$\ts}\fi}
\def\msol {\ifmmode{{\rm M}_{\odot}}\else{M$_{\odot}$\ts}\fi}
\def\lsun {\ifmmode{{\rm L}_{\odot}}\else{L$_{\odot}$\ts}\fi}
\def\cii  {\ifmmode{{\rm [C}{\rm \scriptstyle II}]}\else{[C\ts {\scriptsize II}]\ts}\fi}
\def\ciii  {\ifmmode{{\rm [C}{\rm \scriptstyle III}]}\else{[C\ts {\scriptsize III}]\ts}\fi}
\def\ci   {\ifmmode{{\rm C}{\rm \scriptstyle I}}\else{C\ts {\scriptsize I}\ts}\fi}
\def\m    {\ifmmode{\mu {\rm m}}\else{$\mu$m}\fi}
\def\hi   {\ifmmode{{\rm H}{\rm \scriptstyle I}}\else{H\ts {\scriptsize I}\ts}\fi}
\def\hii  {\ifmmode{{\rm H}{\rm \scriptstyle II}}\else{H\ts {\scriptsize II}\ts}\fi}
\def\nii  {\ifmmode{{\rm [N}{\rm \scriptstyle II}]}\else{[N\ts {\scriptsize II}]\ts}\fi}
\def\oiii {\ifmmode{{\rm [O}{\rm \scriptstyle III}]}\else{[O\ts {\scriptsize III}]\ts}\fi}
\def\hh   {\ifmmode{{\rm H}_2}\else{H$_2$\ts}\fi}
\def\nhh  {\ifmmode{N({\rm H}_2)}\else{$N$(H$_2$)\ts}\fi}
\def\microns {\ifmmode{\mu{\rm m}}\else{$\mu$m\ts}\fi}

\def\barolo {\textsuperscript{3D}\textsc{Barolo} }

\definecolor{lightred}{HTML}{ffd9d9}
\definecolor{lightorange}{HTML}{ffe6cb}
\definecolor{lightyellow}{HTML}{fff6ca}
\definecolor{lightgreen}{HTML}{daffcd}
\definecolor{lightblue}{HTML}{d7e5ff}
\definecolor{lightpurple}{HTML}{f5d9ff}
\definecolor{lightpink}{HTML}{ffcde3}
\definecolor{lightgray}{HTML}{dfdfdf}

\begin{document}

   \title{Structure and kinematics of a massive galaxy at z $\sim$ 7}


 \author{A. C. Posses
          \inst{1} 
          \and
          M. Aravena\inst{1} 
          \and
          J. González-López\inst{1,2}
          \and 
          R. J. Assef \inst{1}
          \and 
          T. Lambert \inst{1}
          \and 
          G. C. Jones \inst{3,4}
          \and
          R. J. Bouwens \inst{5}
          \and
          D. Brisbin \inst{6}
          \and
          T. Díaz-Santos \inst{7,8}
          \and
          R. Herrera-Camus \inst{9}
          \and
          C. Ricci \inst{1,10} 
          \and
          R. Smit \inst{11}
          }

   \institute{N\'ucleo de Astronom\'{\i}a, Facultad de Ingenier\'{\i}a y Ciencias, Universidad Diego Portales, Av. Ej\'ercito 441, Santiago, Chile. \\\email{ana.posses@mail.udp.cl}
   \and
   Las Campanas Observatory, Carnegie Institution of Washington, Casilla 601, La Serena, Chile
   \and
   Cavendish Laboratory/Kavli Institute for Cosmology, University of Cambridge, 19 J. J. Thomson Ave., Cambridge CB3 0HE, UK
   \and
   Department of Physics, University of Oxford, Denys Wilkinson Building, Keble Road, Oxford OX1 3RH, UK
   \and 
   Leiden Observatory, Leiden University, NL-2300 RA Leiden, Netherlands
   \and
   Joint ALMA Observatory, Alonso de Cordova 3107, Vitacura, Santiago, Chile
   \and
   Institute of Astrophysics, Foundation for Research and Technology–Hellas (FORTH), Heraklion, GR-70013, Greece.
   \and 
   School of Sciences, European University Cyprus, Diogenes street, Engomi, 1516 Nicosia, Cyprus
   \and
   Departamento de Astronomía, Universidad de Concepción, Barrio Universitario, Concepción, Chile
   \and 
   Kavli Institute for Astronomy and Astrophysics, Peking University, Beijing 100871, People’s Republic of China
   \and 
   Astrophysics Research Institute, Liverpool John Moores University, 146 Brownlow Hill, Liverpool L3 5RF, UK
   }


   \date{Received XXX; accepted XXX}

 
  \abstract
   {Observations of the rest-frame UV emission of high-redshift galaxies suggest that the early stages of galaxy formation involve disturbed structures. Imaging the cold interstellar medium (ISM) can provide a unique view of the kinematics associated with the assembly of galaxies.}
   {In this paper, we analyzed the spatial distribution and kinematics of the cold ionized gas of the normal star-forming galaxy COS-2987030247 at $z=6.8076$, based on new high-resolution observations of the \cii$158\rm\mu$m line emission obtained with the Atacama Large Millimeter/submillimeter Array (ALMA). }
   {These observations allowed us to compare the spatial distribution and extension of the \cii and rest-frame UV emission, model the \cii line data-cube using \barolo, and measure the \cii luminosity and star formation rate (SFR) surface densities in the galaxy subregions.}
   {The system is found to be composed of a main central source, a fainter north extension, and candidate \cii companions located 10-kpc away.  
   We find similar rest-frame UV and \cii spatial distributions, suggesting that the \cii emission emerges from the star-forming regions. The agreement between the UV and \cii surface brightness radial profiles rules out diffuse, extended \cii emission (often called a \cii halo) in the main galaxy component.
   The \cii velocity map reveals a velocity gradient in the north-south direction suggesting ordered motion, as commonly found in rotating-disk galaxies. But higher resolution observations would be needed to rule out a compact merger scenario. Our model indicates an almost face-on galaxy ($\rm i \sim$ \ang{20}), with an average rotational velocity of 86 $\pm$ 16 km s$^{-1}$ and a low average velocity dispersion, $\sigma <30$ km s$^{-1}$. This result implies a dispersion lower than the expected value from observations and semi-analytic models of high redshift galaxies. Furthermore, our measurements indicate that COS-2987030247  and its individual regions lie systematically within the local $L_{\rm [CII]}-$SFR relationship, yet slightly below the local $\Sigma_{[CII]}$-$\Sigma_{UV}$ relation.}
   {We argue that COS-2987030247 is a candidate rotating disk experiencing a short period of stability which will be possibly perturbed at later times by accreting sources.}


   \maketitle
%
\textbf{}
\section{Introduction}

The early stages of the build-up of galaxies are thought to be marked by gas accretion from the intergalactic medium (IGM) as well as episodes of major galaxy mergers \citep{2005keres, 2006hopkins,2009dekel,2011Bournaud}. These processes provide the fuel for star formation, which  is regulated by stellar and AGN feedback \citep{2000heckman,2008hopkins,2013silk,2015king}. Furthermore, the far ultraviolet (FUV) photons produced by the first stars and early galaxies could ionize the previously neutral IGM. It led to the so-called cosmic reionization, which is believed to have taken place at $z\sim20-6$ \citep{2015robertson}. 

Normal star-forming galaxies in the early universe ($z\sim8-6$), located close to the knee of the UV luminosity function, are expected to be sufficiently abundant and produce enough ionizing photons to play an important role in the process of cosmic reionization \citep{2020dayal,2020naidu}. Thus, describing the physical properties and mechanisms of assembly of these early systems, through morphological and dynamical studies, remains a fundamental step to understand the formation and evolution of galaxies, as well how these sources influenced the evolution of the universe. 

In the last few decades, a large number of studies have taken advantage of the high-angular resolution optical and near-infrared images to investigate the structure of $z>6$ galaxies. The Hubble Space Telescope (HST) images, which at these redshifts trace the rest-frame UV emission, and thus the star-forming component of such galaxies, have mostly shown (multi-component) clumpy structures, likely indicating an early stage of mass assembly \citep{2015capak, 2018carniania, 2018carnianib, 2017bowler, 2017matthee, 2019matthee, 2020lefevre}. The rest-frame UV radiation, however, might be affected by dust obscuration, which can lead to patchy geometry, and then preventing us from acquiring a reliable picture of the galaxy structures. Observations of the different components of galaxies, including the star formation, stars and the cold interstellar medium (ISM) at wavelengths not affected by dust obscuration, are thus necessary to provide a full description of the galaxy structures and thereby mechanisms for galaxy growth \citep{2020forster}.  

The Atacama Large Millimeter/submillimeter Array (ALMA), due to its great sensitivity and angular resolution, has opened a new window on the distribution of the cold ISM by exploring the far-infrared regime through dust continuum and fine-structure emission lines. In particular, the $^2$P$_{3/2}-^{2}$P$_{1/2}$ transition of the singly ionized carbon atom (hereafter \cii; centered at 158$\mu$m) is one of the brightest cooling lines\footnote{Due to its low ionization potential of 11.3 eV} in the far-infrared regime \citep{1991stacey,2010stacey} and a tracer of multiple gaseous phases \citep[cold molecular, neutral and ionized gas associated with photon-dissociation regions;][]{2003wolfire,2015vallini,2019clark}. 

At $z > 2$, the \cii emission line shifts into the (sub)millimeter atmospheric windows, which are widely accessible from the ground, and thus  has become a flagship of ISM studies of early galaxies. In recent years, a plethora of \cii line detections have been reported \citep[e.g.][]{2015capak,2015willott,2018harikane, 2018smit, 2018carniania}, including those by the ALPINE \citep{2020lefevre} and REBELS \citep{2021bouwens} ALMA large programs, particularly designed to provide \cii line measurements for normal star-forming galaxies at $z=4-6$ and $z=6-9$, respectively. While the number of \cii detections have increased rapidly (mostly at $\sim1''$ resolution), only a few observations to date have been able to resolve the \cii emission at $z > 4$ to enable the characterization of the galaxy's structures and kinematics \citep{2020neeleman,2020rizzo, 2021lelli,2021herrera-camus,2021Rizzo}.    

The wealth of relatively low-resolution (i.e., $\sim1\arcsec$) observations have shown complex \cii structures, revealing  clumpy/irregular in some cases  \citep[]{2016inoue,2017matthee,2018carniania, 2018carnianib}, as well as regular velocity fields,  in others \citep[Fig. 3 in][]{2018smit, 2020neeleman}. In some systems, the brightest regions in \cii and UV emission do not spatially match, while in others the presence of \cii (UV) clumps with no UV (\cii) counterparts has been reported \citep{2015capak, 2020fujimoto}. In a similar vein, several objects display extended \cii emission when compared to the UV component. It is still unclear what is causing this difference, but some possible explanations include differential dust obscuration, outflows and/or infalling satellites \citep{2019fujimoto, 2020fujimoto}. 


In this paper, we report sensitive ALMA high-resolution \cii line observations of the typical star-forming galaxy COS-2987030247 (hereafter as COS2987) at $z=6.8076$. This system was a $z\sim7$ Lyman-break galaxy candidate with a high [OIII]+H$\beta$ equivalent-width (EW) obtained from optical to infrared spectral energy distribution (SED) modelling \citep{2015smit, 2017laporte}, and later confirmed to be a gas-rich star-forming galaxy based on the detection of its bright \cii line with ALMA \citep{2018smit}. Dynamical analysis of the \cii line velocity field in low-resolution observations ($1.1''$x$0.7''$, translated to 5.8×3.7 kpc at z = 6.8) suggested that disk rotation could be in place \citep{2018smit}. Here, we analyze in detail the morpho-kinematic properties of this galaxy and its close environment. 

In Section \ref{sec:data}, we describe the general properties of our target, as well as the ALMA observations. In Section \ref{sec:results}, we analyze the structure and spatial extension of the \cii and UV emission, and we present a dynamical analysis. In Section \ref{sec:discussion}, we discuss the ISM properties derived from  resolved measurements of the star formation rate (SFR) and \cii surface densities. We discuss the implications of our results in the context of cosmic evolution of the intrinsic velocity dispersion of rotating disks. In Section \ref{sec:conclusions}, we list the main results. We assume a standard $\Lambda$CDM cosmology with: $\Omega_{\Lambda} = $0.7, $\Omega_{M} = $0.3, H$_{0} = 70$ km s$^{-1}$ Mpc$^{-1}$ throughout the article, which leads to a physical conversion of 5.312 kpc/\arcsec.

\section{Data}
\label{sec:data}
\subsection{Target}

COS2987 is a star-forming galaxy located at $z = 6.8076 \pm 0.0002$ \citep{2018smit}. It was initially detected as a Lyman-break galaxy by \citet{2015smit} in the CANDELS/COSMOS field \citep{2007scoville,2011Koekemoer}. Rest-frame UV to far-infrared observations, as well as SED modeling indicate a massive star-forming galaxy, located within the expected main-sequence of star formation at this redshift, and low dust content due to non-detection of dust continuum emission \citep{2018smit}. The multi-wavelength SED (including 4.5$\mu$m Spitzer/IRAC band) yields a stellar mass of  $1.7^{+0.5}_{-0.2}\times10^{9}{\rm M}_{\odot}$ \citep{2018smit}. Measurements of Lyman-$\alpha$ line emission indicates a broad and strong emission compared to other sources at the same redshift \citep{2017laporte}. 
. 

\citet{2018smit} identified tentative evidence that the galaxy is a rotating system based on the detection of the \cii line emission using ALMA observations at $\sim0.8''$ resolution (PI: Smit, ALMA ID: 2015.1.01111.S). This finding makes the galaxy a unique target for confirmation of its dynamical state, given that most systems at $z>5$ showed clumpy irregular morphology \citep{2015capak}. The main physical properties of COS2987 from the literature as well as those derived from this work are listed in Table ~\ref{tab:properties-central}. 

\subsection{ALMA observations}

ALMA observations were obtained during Cycle 5 in the C43-5 configuration with 43 antennas, (PI: Aravena, PID: 2018.1.01359.S) with 3.6 hours on-source integration time. The ALMA band 6 receivers were used to target the redshifted \cii line emission at 243.3465 GHz. Observations were taken with 4 spectral windows (SPWs), each with a total bandwidth of 1.875 MHz, and a native channel resolution of 7.813 MHz ($\approx$ 9.6 km s$^{-1}$). One of the SPWs was centered on the \cii emission line, and the rest of the SPWs were used to measure the underlying continuum emission. 


We combined the new ALMA Cycle 5 data with previous lower resolution observations from the original \cii detection (PID: 2015.1.01111.S; Smith et al. 2018).  These observations spent 24 minutes of integration on-source. These data adds important information of shorter baselines, which are sensitive to extended emission. The lower and higher resolution data sets were calibrated with the Common Astronomy Software Applications package pipeline \citep[CASA;][]{2007mcmullin} using versions 4.5.3 and 5.4, respectively. The quasar J0948+0022 was used as the phase calibrator for both observations, while J1058+0133 and J0854+2006 were chosen as amplitude calibrators for the previous and latest data sets, respectively. No additional flagging was deemed necessary given the good quality of the observations.

We employed the \textit{tclean} task to generate a continuum image and a data cube for the SPW containing the \cii emission line, using natural weighting to preserve sensitivity. To obtain a continuum image, we used all the channels in each SPW, excluding the ones within the velocity range [-250, 250] km s$^{-1}$, centered at the line. For the cube, we binned the data to a channel resolution of 30 km s$^{-1}$, which is a compromise between sensitivity and velocity resolution. The cleaning was done interactively until no significant emission was left in the residual. Since no continuum emission was detected, we did not perform a continuum subtraction in the uv plane. This procedure yielded a noise level of 6 $\mu$Jy beam$^{-1}$ and 0.134 mJy beam$^{-1}$ for the continuum image and line cube, respectively, and a synthesized beam size of 0.44$\arcsec\times0.35\arcsec$ (PA$=50\deg$; natural weighting). This scale translates into a physical size of $2.3\times1.8$ kpc$^2$, at z = 6.8. The angular resolution obtained for previous observations of this galaxy is $\sim$2 times coarser: 1.07$\arcsec\times0.72\arcsec$, corresponding to $5.7\times3.82$ kpc$^2$ (Appendix \ref{appendix:data-products}).

\subsection{Ancillary data}

We used the F160W-band image ($\rm\lambda_{rest-frame} \sim$ 2050 \r{A}) obtained with the Wide Field Camera 3 (WFC3) of the HST from the CANDELS survey (PI: Sandra Faber, ID: 12440). It yields a view of the rest-frame near-UV emission at the redshift of the source. Rest-frame UV images are also available from the F125W-band \citep{2011Koekemoer}. We chose the F160W-band for the spatial analysis, mostly for consistency with the analysis conducted by \citet{2018smit} . The F125W-band was employed to measure the SFR, since it represents a rest-frame far-UV regime ($\lambda_{rf} \sim 1600$ \r{A}). Both F160W and F125W images have a point spread function (PSF) of 0.18$\arcsec$, a pixel scale of 0.06$\arcsec$/pix and have been astrometrically matched to the Gaia DR2 release \citep{2018gaiadr2}.

\begin{table}
\caption{Observational and physical properties of COS2987}
\centering          
\begin{tabular}{l c c}     
\hline\hline       
Property & Value & Reference \\ 
\hline                    
   Right Ascension                           & +10:00:29.86            & 2 \\  
   Declination                               & +02:13:02.19            & 2 \\ 
   $z_{\mathrm{[CII]}}$                      & 6.8076 $\pm$ 0.0002     & 1 \\ 
   SFR$_{UV}$ (M$_{\odot}$ yr$^{-1}$)        & 16 $\pm$ 9.0            & 2 \\ 
 
   SFR$_{IR}$ $^{a}$ (M$_{\odot}$ yr$^{-1}$) & $\leq$ 1.7              & 2 \\ 
   Stellar mass ($10^{9}$ M$_{\odot}$)       & 1.7$^{+0.5}_{-0.2} $    & 2 \\ 
   $\beta_{UV}$                              & -1.18 $\pm$ 0.53        & 1 \\ 
   
   \\
   S/N (\cii)                                & 8.4$\sigma$             & 2\\ 
   
   \cii line flux (Jy km s$^{-1}$)           &  0.15 $\pm$ 0.02        & 2 \\ 
   FWHM$_{\rm [CII]}$ (km s$^{-1}$)          &  126 $\pm$ 16           & 2  \\ 
   $S_{\rm 158\mu m}$ $^{a}$ ($\mu$Jy)       &    $\leq$ 18            & 2  \\ 
   L$_{\rm [CII]}$ ($10^{8}$ L$_{\odot}$)    & 1.6 $\pm$ 0.2           & 2 \\ 
   L$_{\rm FIR}$ ($10^{10}$ L$_{\odot}$)     & $\leq$1.15              & 2 \\ 
   \\ 
   r$_{\rm {UV}}^{eff}$ $^{b}$               & 0.31"$^{+0.06}_{-0.05}$ & 2 \\ 
   r$_{\rm [CII]}^{eff}$ $^{b}$               & 0.38"$^{+0.02}_{-0.02}$ & 2\\ 
   \\
   v$_{\rm rot}$ (km s$^{-1}$)              & 86 $\pm$ 16               & 2 \\  
   v$_{\rm disp}$ (km s$^{-1}$)             & $\leq$ 30                & 2 \\ 
   Conversion (kpc/$"$)                     & 5.312                    &  \\ 

\hline  
\hline

\end{tabular}
\flushleft{{\bf Notes:} The values derived in this work were obtained in the region containing the main central source C with a 0.5\arcsec-aperture, see Section \ref{sec:spatial-distribution}. $^{a}$ 3-$\sigma$ upper limit  ,$^{b}$  Considering an exponential radial profile (S\'ersic index = 1.0); References: [1] \citet{2018smit} and [2] This work. } 
\label{tab:properties-central}  
\end{table}

\section{Results}
\label{sec:results}
\subsection{Spatial distribution}
\label{sec:spatial-distribution}

\label{section-spatial}

\begin{figure*}[ht]
    \centering
    \includegraphics[width=\textwidth]{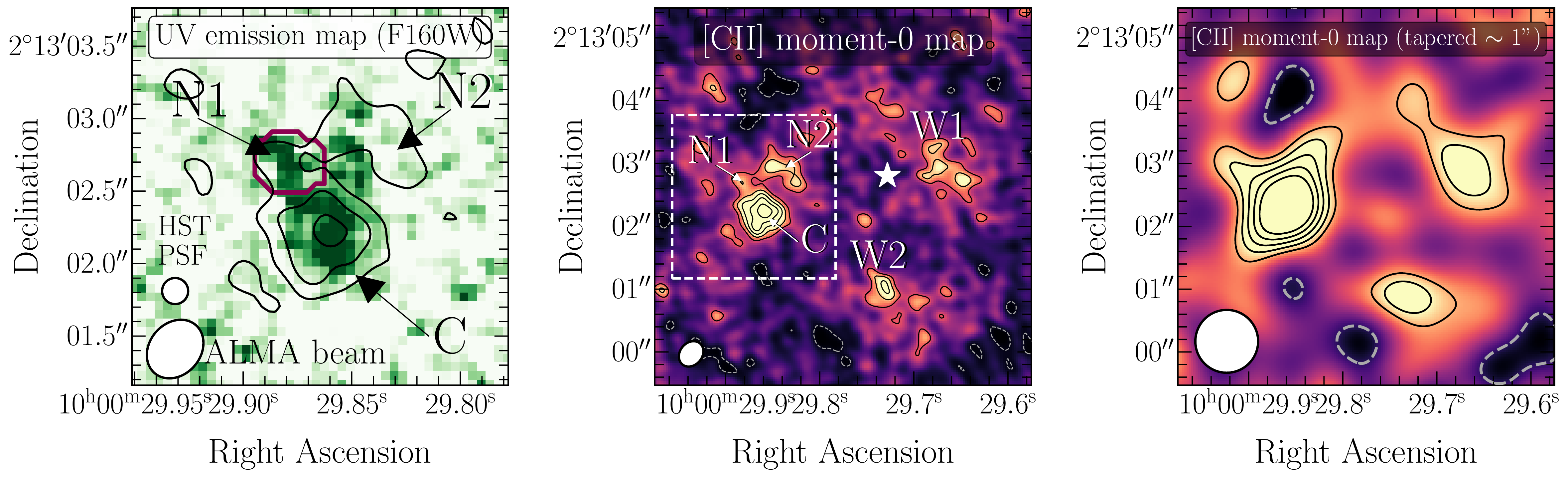}
    \caption{The rest-frame UV and \cii line maps of the  galaxy COS2987 and its surroundings. \textbf{(a) Left:} The HST F160W-band image in the background shows the rest-UV emission in a $1.3\arcsec\times1.3\arcsec$ region around the galaxy. The \cii integrated map (moment-0) is overlaid as black contours shown at 2-, 4- and 6-$\sigma$, where $\sigma$ is the rms noise level of the moment-0 map. The region containing the foreground galaxy identified by \citet{2017laporte} is highlighted by the purple line. \textbf{(b) Middle:} Zoom-out version of the \cii moment-0 map toward COS2987 (6\arcsec x 6\arcsec). The region shown in the left panel is represented by the dashed white line. We marked the locations of the central source (C), north-east emission (N1), north-west arm (N2), west emission (W1) and south-west emission (W2). The black contours represent the integrated \cii line emission at 2-, 3-, 4-, 5- and 6-$\sigma$ levels and the dashed white contour represent the -2-$\sigma$ level.  The white star corresponds to a galaxy observed in th F160W-band map at z$_{phot} = 1.73$ \citep{2016laigle}.\textbf{(c) Right:} \cii line integrated map tapered to a 1$\arcsec$ beam size, showing the same region as in the central panel. The black contours represent the 2-, 4-, 6-, 8- and 10-$\sigma$ levels and the dashed white contour represent the -2-$\sigma$ level.}
    \label{fig:UV-[CII]}
\end{figure*}

\begin{figure}[ht]
    \centering
    \includegraphics[scale=0.42]{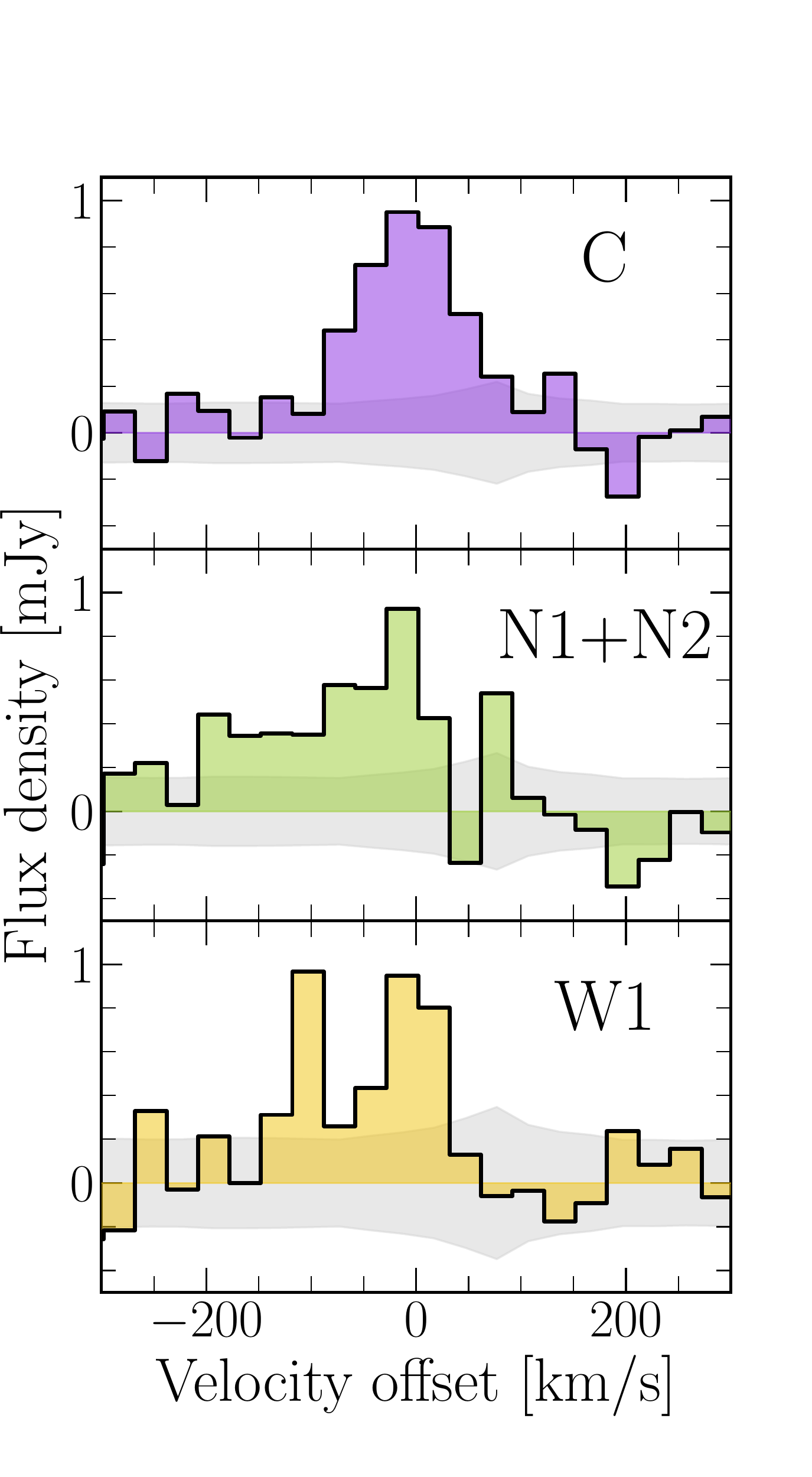}
    \caption{Spectra ranging from -300 to 300 km s$^{-1}$ centered at the \cii observed-frame frequency at z = 6.8076 (243.3465 GHz), based on a channel resolution of 30 km s$^{-1}$. The gray region corresponds to the rms per channel.  The spectra represent the: \textbf{(a) Upper panel:} central source C, \textbf{(b) Central panel:} N1+N2 emissions and \textbf{(c) Bottom panel:} West emission. The regions are marked in Figure~\ref{fig:UV-[CII]}. }
    \label{fig:UV-[CII]-spectra}
\end{figure}

We analyzed the spatial distribution of the  rest-frame UV emission, traced by the HST F160W-band image, and the \cii line emission from ALMA, in the left panel of Figure \ref{fig:UV-[CII]}. We find a good match between the spatial distribution of both emissions, particularly in the central region where most of the \cii emission is detected. There is a small offset of 0.118$\arcsec$ (corresponding to 0.628 kpc) between the centroids of UV and \cii emissions, which is less than the HST PSF and the ALMA beam size.   Hence, we do not find a divergent location of the UV and \cii emission, as it has been observed in other objects at similar redshift. We refer to this central region as the `C' component in the left and central panel of Figure \ref{fig:UV-[CII]}. A 0.5\arcsec-circular aperture photometry reveals a \cii line flux of $0.15 \pm 0.02$ Jy km s$^{-1}$, which corresponds to an 8.4$\sigma$ detection.

To the north of C, we find two clumpy structures traced by both the UV and \cii emissions. One of them extends to the north-east, referred as `N1', and the other resembles an arm-shaped structure to the north-west, labelled as `N2'. Component N1 is located 0.60$\arcsec$ north-east from C (measured at the centroids of C and N1 in the \cii image), while N2 is located at 0.73$\arcsec$ north-west from C. This corresponds to a projected physical distance of 3.2 and 3.9 kpc for N1 and N2, respectively. Given the moderate significance of each of these structures, we performed aperture photometry to measure their individual \cii fluxes. We find that N1 and N2 have integrated \cii line fluxes of 0.02 $\pm$ 0.01 Jy km s$^{-1}$ and 0.07 $\pm$ 0.02 Jy km s$^{-1}$, yielding significances of 2.2$\sigma$  and 4.1$\sigma$,  respectively (Appendix \ref{sec:appendix-sourceproperties}).  We highlight the fact that these components were not seen by \citet{2018smit}, mainly because of the lower depth, and were likely blended with the main source.

\citet{2017laporte} indicated that the rest-UV source associated with N1 corresponds to a foreground galaxy at z$\rm_{spec}$ = 2.099. Based on this, and given that the candidate \cii emission is of low significance, we consider it as a foreground object. Thus, for further analysis, we mask out the region within N1, which corresponds to the purple contour in the left panel of Fig.~\ref{fig:UV-[CII]}. In the case of N2, the significance of the \cii line emission and its extended structure suggest that this is associated with the main galaxy C at z $\sim$ 7. The F160W-band image also shows an extended UV emission in the north-west of the central galaxy C, but not perfectly colocated to N2.

To check for the existence of other emitting regions at the redshift of the main galaxy, we searched for \cii emitters in the wider field of the ALMA \cii map (Fig.~\ref{fig:UV-[CII]} - center). Two other sources are identified as candidate \cii line emitters to the west and south-west of COS2987 (labelled as W1 and W2, respectively). They are located at 15 kpc and 12 kpc from the central source C, with flux densities of  0.10 $\pm$ 0.03 Jy km s$^{-1}$ ($\sim$ 5.2$\sigma$) and 0.06 $\pm$ 0.02 Jy km s$^{-1}$ ($\sim$ 3.6$\sigma$), respectively. We only found a UV counterpart in the F160W image (Appendix \ref{appendix:data-products}) to source W1, located at $0.95''$ from the peak \cii emission and marked as a white star in Fig. ~\ref{fig:UV-[CII]} (central panel). It is consistent with a photometric redshift of 1.73 based on the COSMOS2015 catalog \citep{2016laigle}. The lack of counterpart for W2 supports the fact that the candidate \cii positive signal is spurious. Furthermore, the significant distance between the UV counterpart to W1, and the conflicting photometric redshift of this source imply either that this is not the actual counterpart, in which case the \cii emission might be real but without a UV counterpart, or that the candidate \cii line is a noise feature. The spectra centered at the \cii observed frequency of C, N1+N2 and W1 are shown in Figure~\ref{fig:UV-[CII]-spectra}.

A 1\arcsec-tapered map (to maximize the extended emission significance of the emitters; Fig. \ref{fig:UV-[CII]} - right) reveals a tentative detection of the clumps, but the low significance of W2 and N2 require deeper observations to confirm or discard the sources. Finally, we performed two extra tests to check the reliability of the candidates. Firstly, we split the visibility data in two different groups of observing scans (both containing scheduling blocks from the previous and latest data), and recreated the \cii cubes and maps. We could recover sources N2 and W1, but not W2 in both maps, which suggest that the latter is likely a product of a noise pattern. Secondly, we explored the moment-0 maps with different weightings and taperings to exclude the possibility that the 4 candidates are artifacts generated by the side-lobes of the beam. The candidates become fainter in the moment-0 maps, particularly when we use uniform parameters below 1. Nevertheless, as we combine the standard Briggs weighting with a uniform parameter of 0.5, and taper the visibilities, the N2 and W1 candidates still remain with moderately significance. Deeper observations will be key to ultimately confirm or reject the reality of these surrounding candidates sources.

\subsection{Radial profile and sizes}

\begin{figure}
    \centering
    \includegraphics[scale=0.55]{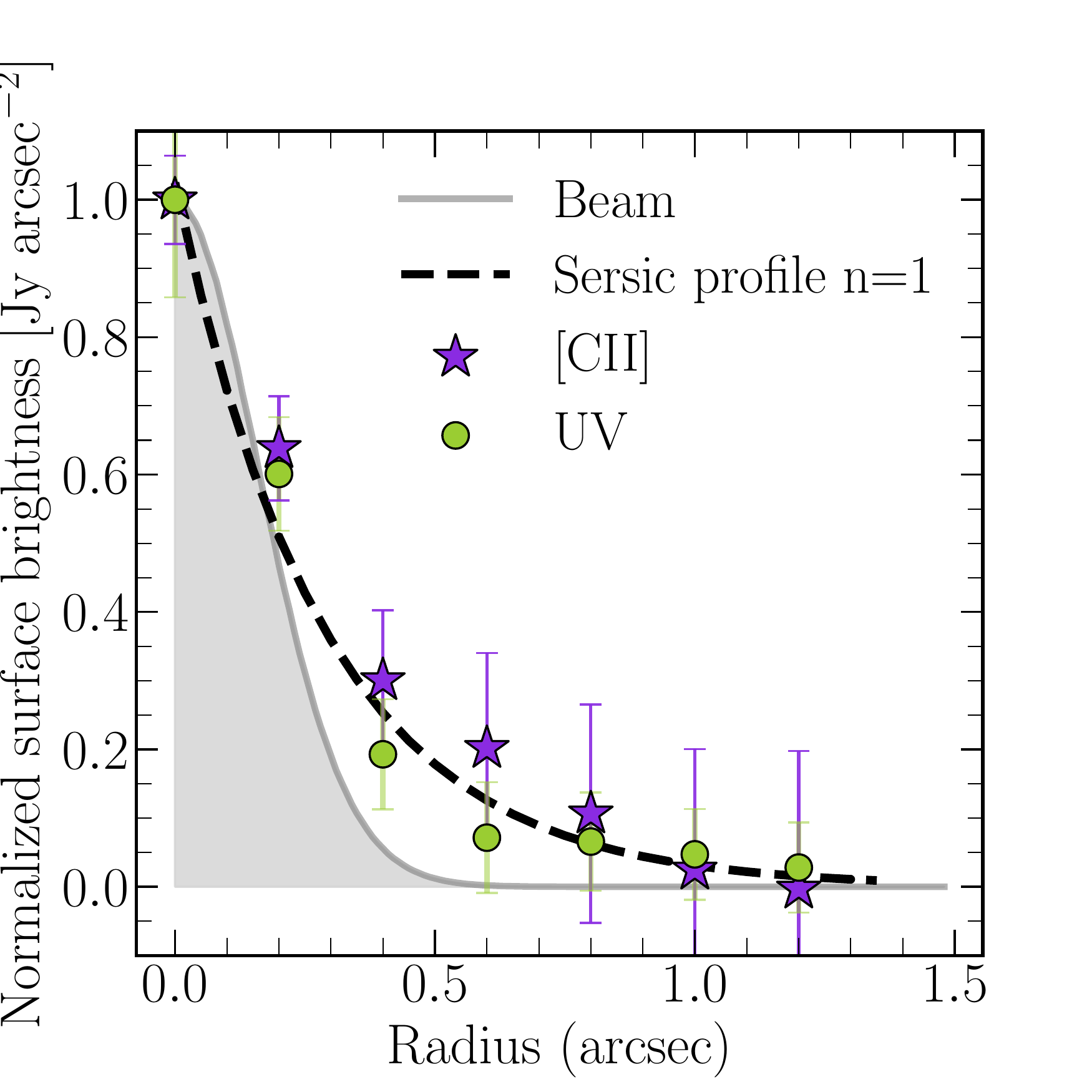}
    \caption{Radial profile of the surface brightness of the UV continuum (green circles) and \cii line (purple stars). Each map is convolved with the PSF/beam of the other image to match the resolution of $0.47\arcsec\times0.40\arcsec$. There is no detection of a dust continuum, therefore no radial profile is shown. As a reference, we added the radial profile of the convolved beam and a exponential profile (S\'ersic index = 1) with a r$_{eff}$ = 0.38\arcsec as a solid gray line and dashed black line, respectively.}
    \label{fig:SB-profile}
\end{figure}

We aimed to quantify whether the ISM component traced by the \cii line emission has a similar spatial distribution to the star-forming component traced by the rest-UV emission. Thus, we measured their effective sizes and compared their radial profiles. We first fitted a two-dimensional S\'ersic profile \citep{1963sersic} to the \cii brightness distribution in the moment-0 map and UV map using a Monte Carlo Markov Chain (MCMC) approach (Appendix \ref{sec:appendixsersic}). The S\'ersic profile is mainly composed of two parameters: the S\'ersic index, $n$, which describes the curvature of the profile, and the effective radius, $r_{\rm eff}$ defined as the radius at which half of the total brightness was emitted. We convolved the intrinsic and not inclination-corrected model with the synthesized beam before comparing to the moment-0 map. We obtained a consistent radial profile distribution with $n_{\rm [CII]}=1.62^{+0.59}_{-0.43}$ and $r_{\rm eff,[CII]}=0.47$\arcsec$^{+0.12}_{-0.07}$($\sim$ 2.3 kpc), but with a largely unconstrained $n_{\rm UV}$, possibly due to the faint data. When we set $n = 1$, we found a compatible effective radius for both components: $\rm r_{ eff,[CII]}$ = $0.38$\arcsec$^{+0.02}_{-0.02}$ ($\sim$ 2 kpc) and $r_{\rm eff,F160W} = 0.31$\arcsec$^{+0.06}_{-0.05}$ ($\sim$ 1.6 kpc). We stick in the further analysis with n = 1 in order to be consistent to other analysis in the literature \citep{2020fujimoto}



COS2987 has one clear component extending to the north-west (N2), as suggested by both the \cii and rest-UV maps. This prompted us to compare the extension of the \cii and rest-UV emission in the source (C + N2), and thus check for an extended \cii halo. We measured the radial surface brightness profile in concentric rings as a function of radius. This procedure is not available for the dust component of the galaxy, since there is no detection of the continuum emission.

Since the ALMA \cii and F160W rest-UV maps have different angular resolutions, we followed the procedure described in \citet{2020fujimoto}. We convolved the ALMA (HST/F160W) image with the HST/F160W (ALMA) PSF, generating a resolution of $0.47\arcsec \times0.40\arcsec$, which translates into $2.5\times2.1$ kpc$^2$ in physical size. For each image, we computed the surface brightness measuring the \cii and rest-UV flux in $0.2\arcsec$-width rings, covering the region from the central peak to the end of the N2 structure. As stated in the previous Section, there is a foreground galaxy located to the  north-east (N1) of the central source. Thus, we masked the pixels corresponding to this structure in both the ALMA and HST/F160W images. However, given its faintness, including or not this region shows no significant difference in the resulting radial profile.

The \cii and rest-UV emission radial profiles are shown in Fig. \ref{fig:SB-profile}, and they are in good agreement within the uncertainties. There is a slight increase in \cii emission for the region closer to the north arm N2 ($r\sim0.6"$), however it does not represent a substantial difference compared to the rest-UV. We find that the  \cii emission in this case closely follows the rest-UV emission, in shape and size. Since we do not recognize major differences between the \cii and rest-UV profiles, we conclude that COS2987 does not contain a particularly extended \cii emission or a \cii halo.


\subsection{Dynamical analysis}

\begin{figure*}[ht]
    \makebox[\textwidth][c]{\includegraphics[width=1.2\textwidth]{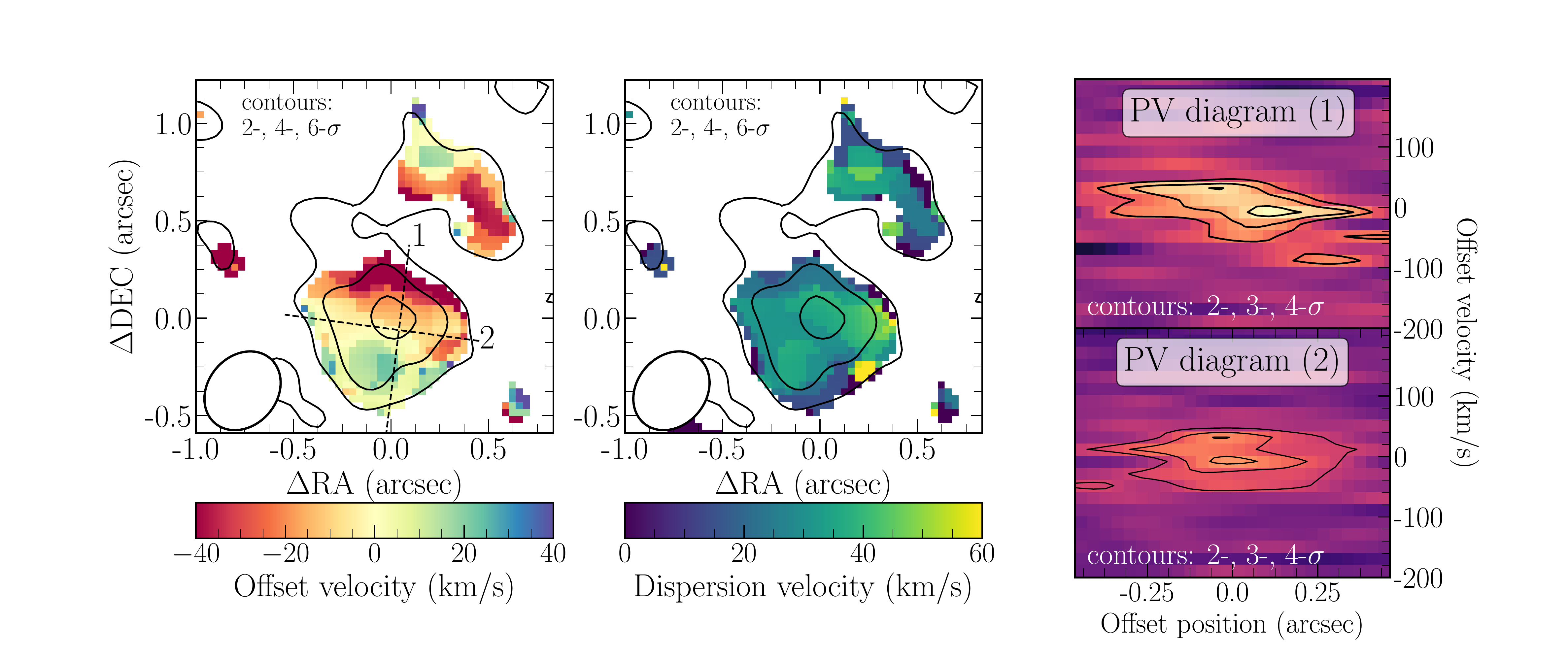}}
    \caption{ The velocity (Moment-1) and dispersion (Moment-2) maps of the galaxy COS2987 (with respect to the observed frequency of \cii at redshift $z = 6.8076$) are shown in the left and right panels, respectively. The black contours correspond to the 2-, 4-, 6-$\sigma$ levels of the \cii integrated (moment-0) map. In the right panel, we plot the position-velocity diagram of the slits 1 and 2 shown in the left panel by dashed lines. The black contours correspond to the 2-, 3-, and 4-$\sigma$ levels of the position-velocity diagram.}
    \label{fig:moment-1-2}
\end{figure*}

\begin{figure*}[ht]
    \centering
    \includegraphics[width=\textwidth]{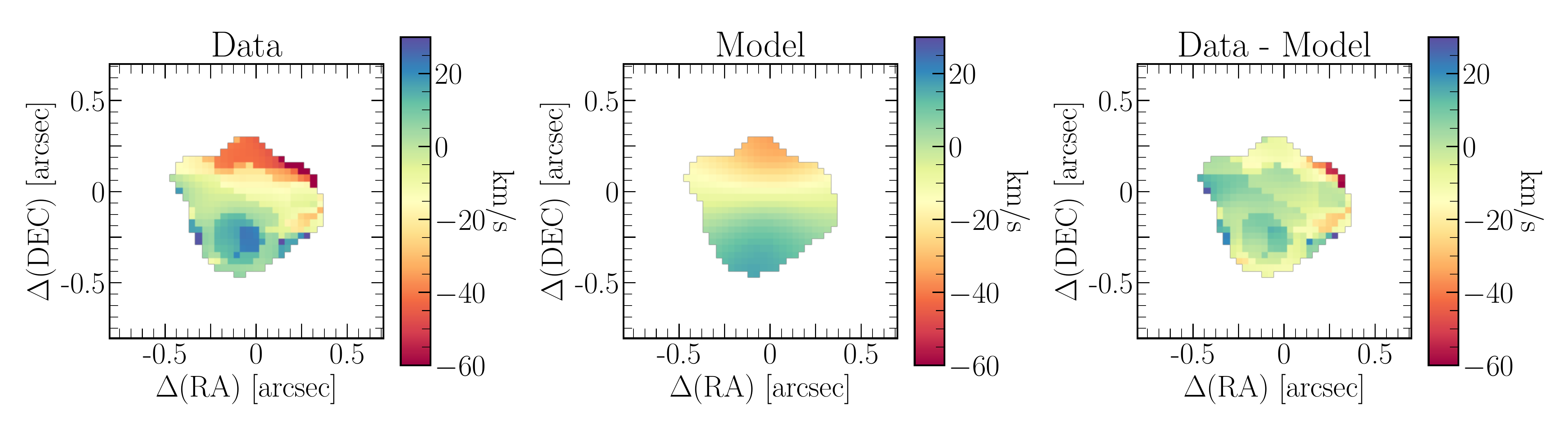}
    \caption{Rotational velocity products of the tilted-ring model fitting obtained using \barolo. The maps are shown in the velocity range of [-60,30] km s$^{-1}$ with respect to the observed frequency of the \cii line at $z = 6.8076$ \textbf{(a) Left:} Observed velocity map of COS2987. \textbf{(b) Middle:} Velocity map of the best-fit model with a inclination of $i = 23\deg$. \textbf{(c) Right:} Residuals of the velocity map.}
    \label{fig:3dbarolorvelocity}
\end{figure*}

\begin{figure*}[ht]
    \centering
    \includegraphics[width=\textwidth]{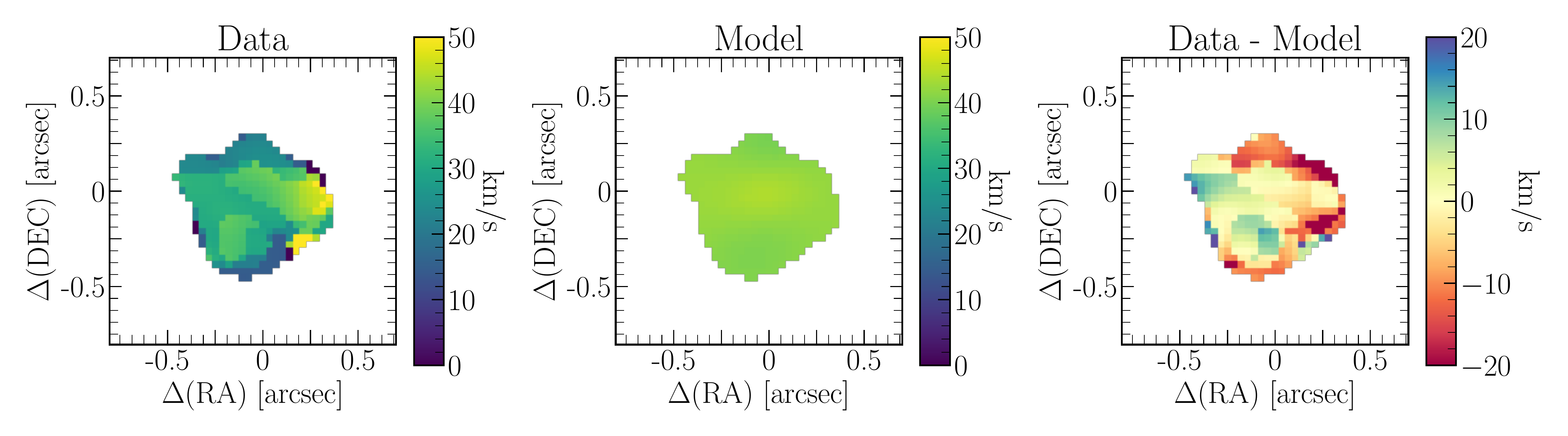}
    \caption{Dispersion velocity products of the tilted-ring model fitting obtained using \barolo. The left and central maps are shown in the velocity range of [0,50] km s$^{-1}$ and the left map in the velocity range of [-20,20] km s$^{-1}$, with respect to the observed frequency of the \cii line at $z = 6.8076$ \textbf{(a) Left:} Observed dispersion velocity map of COS2987. \textbf{(b) Middle:} Dispersion velocity map of the best-fit model with a inclination of $i = 23\deg$. \textbf{(c) Right:} Residuals of the dispersion velocity map.}
    \label{fig:3dbarolodispersion}
\end{figure*}

Based on low-resolution ALMA \cii line imaging, \citet{2018smit} identified a tentative velocity gradient suggesting that a rotating disk is in place in this galaxy. Thus, we analyzed the current dynamical state of COS2987 based on our higher-resolution data. Figure~\ref{fig:moment-1-2} shows the velocity field and dispersion maps (in the left and central panels, respectively) of the central source (C) and north arm (N2).  These maps were generated considering the pixels with fluxes above 2 $\times$ the rms per channel from a cube with a channel resolution of 30 km s$^{-1}$. We also only display pixels with a \cii emission significance greater than 2.5-$\sigma$ in the moment-0 map. The N2 structure does not show a clear trend, due to its fainter nature. Conversely, we confirmed a north-south velocity gradient across the central source (C). The velocity dispersion map is fairly homogeneous, with a low mean dispersion of 29 km s$^{-1}$ and a scatter of 12 km s$^{-1}$. The position-velocity diagram along the major axis of the velocity field (top right panel of Fig.~\ref{fig:moment-1-2}) suggest a tentative pattern typical of  rotating disks, as an increase of the radial velocity with the distance to the center of the galaxy and a flattening after a certain radius. However, this can not be confirmed due to the limited sensitivity of the observations. 

We performed dynamical modeling of our data cube with the \barolo code \citep{2015diteodoro}.  If we assume that COS2987 has a rotating gaseous disk, the modeling retrieves the kinematic parameters of the galaxy, such as the rotational and dispersion velocities. We discuss some of the scenarios that could lead to a gradient in the velocity map in Section~\ref{discussion-dynamics}.  

\barolo is based on a tilted-ring model in the case of a galaxy disk. It divides the model in a given number of concentric rings in order to provide a fit to the observed data cube, taking into account the three spatial and three velocity dimensions. To construct the model, initial guesses of all the parameters need to be provided: the disk position angle, the central position of the disk, the systemic velocity of the system, the radial and rotational velocity, the dispersion velocity, the inclination, and thickness of the disk. This disk model is then convolved to the original data beam size so that the degraded product can reproduce the real data; and thus the code is able to deal with effects due to the resolution, such as beam smearing. The initial iterations begin in the inner annuli and if the parameters reproduce well the data, the code steps to the next annuli. Otherwise, it updates the parameters until it finds the best fit.  

We applied this modelling to the \cii data cube. Using channel resolution higher than 30 km s$^{-1}$, the performance of the fitting routine failed due to the low signal-to-noise per channel. We considered as a detection per channel any emission with intensity equals to or greater than 0.2 mJy beam$^{-1}$ ($2 \times$ the mean rms per channel). We also masked the region to consider pixels with significances above 2.5 $\times$ the rms of the moment-0 map.  We fixed the central position (x$_{0}$,y$_{0}$) on the centroid of the emission in the moment-0 map. The moment 1 map indicates a velocity gradient tilted by nearly $180\deg$ from the negative y-axis, so we set this value as an initial guess position angle, and we allowed it to vary between $170-190\deg$. We let the systemic velocity as a free parameter, since the peak of the line emission seems to be offset by $\sim$ -10 km s$^{-1}$ from the \cii observed frequency. We also fixed the radial velocity at 0 km s$^{-1}$ since the depth and resolution can not allow the detection of outflows from stellar or AGN feedback. As a final step, we let the code run in 3 concentric rings, with a width of $0.18''$. It satisfactorily covers the central source and excludes the north-arm region.

As often pointed out by previous studies \citep{2020rizzo,2021fraternali}, the initial inclination has a crucial influence on the best-fit model. Therefore, we performed trial-and-error tests with initial inclination guesses ranging from $0\deg$ to $90\deg$ and calculated the residuals (|data - model|, weighted by the moment-0 map) in the velocity and dispersion maps. The residuals are lowest when we consider low inclination values (Appendix \ref{appendix:trial-and-error}).  The galaxy has a close to round shape, which likely means that it is close to a face-on configuration. Using these criteria, the galaxy agrees with a system tilted by $\sim15\deg$. So we set as initial guess an inclination of $23\deg$ based on the minimized value found from the velocity and dispersion map. 

In Figures \ref{fig:3dbarolorvelocity} and \ref{fig:3dbarolodispersion}, we show the rotational and dispersion products for the best-fit model. We recovered an average rotational and dispersion velocity of 86 $\pm$ 16 km s$^{-1}$ and 26 $\pm$ 16 km s$^{-1}$, respectively. However,  we preferred to consider in the following analysis the average velocity dispersion as an upper limit of 30 km s$^{-1}$ since the data cube channel resolution is greater than the value found by the code. \citet{2018smit} identified an observed velocity difference of $54\pm20$ km s$^{-1}$, which is tentatively lower than the average rotational velocity found here. However, such value could be strongly affected by the blending of the central source (C) and north arm (N2) due to the lower resolution of the data.

\begin{figure*}
    \centering
    \includegraphics[scale=0.5]{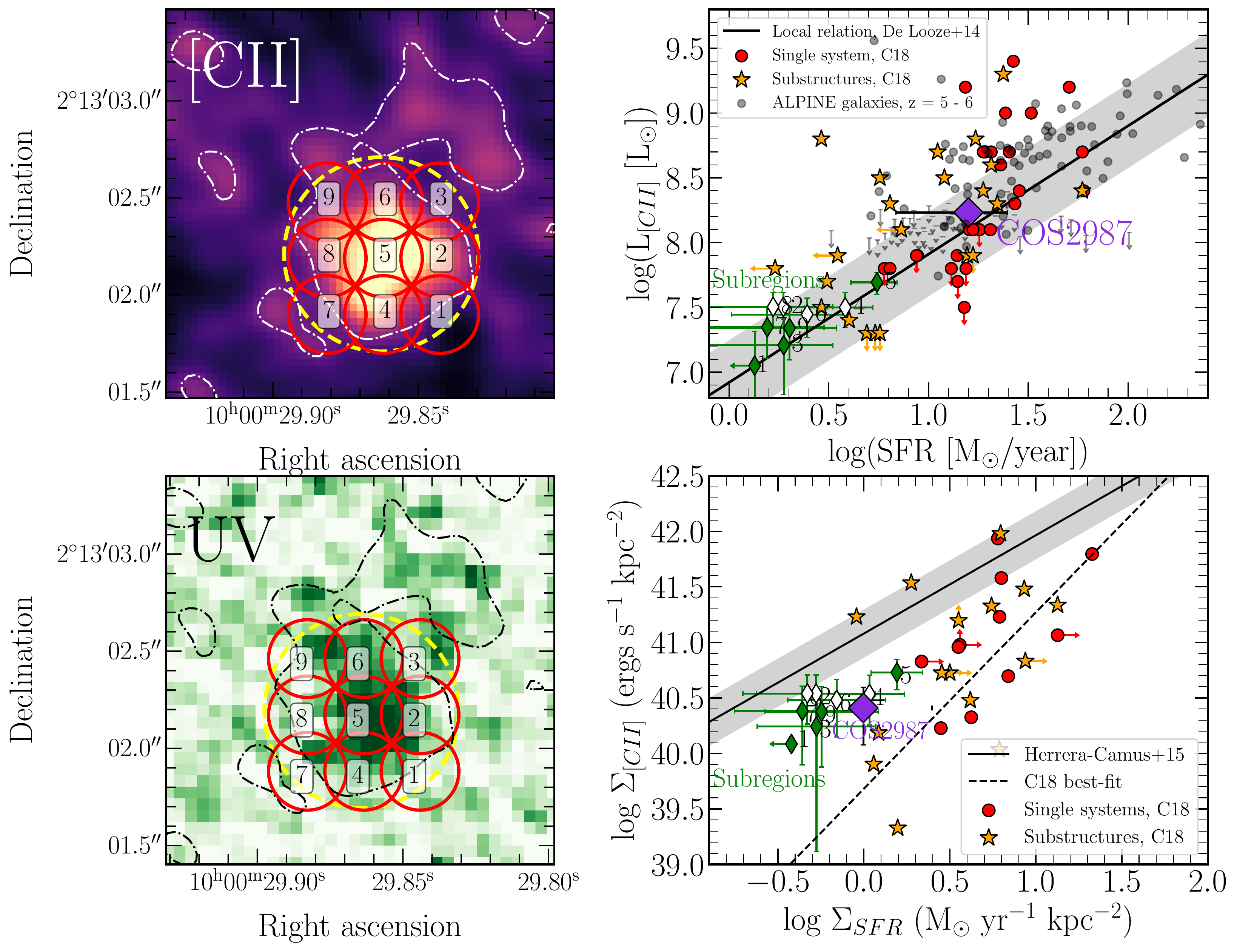}
    \caption{L$_{CII}$-SFR and $\Sigma_{[CII]}$-$\Sigma_{SFR}$ relations for the galaxy COS2987 and its minor components. \textbf{(a) Upper left panel:} \cii moment-0 map with the regions considered as the whole system of COS2987 and its substructures highlighted as a yellow dashed circle and red circles, respectively. The regions within COS2987 are number as displayed in the figure. The dashed-dotted white line shows the 2-$\sigma$ significance level of the galaxy. \textbf{(b) Bottom left panel:} F125W-band map with the same regions displayed in the upper left panel. The 2-$\sigma$ significance level of the \cii moment-0  map is shown as a black dashed-dotted line. \textbf{(c) Upper right panel:} L$_{CII}$-SFR relation. The single system of COS2987 and its substructures are displaced as a purple diamond and thin green/white diamond, respectively. As a reference, the single system and their corresponding substructures of galaxies at z = 5-6 are represented as red circles and orange stars, similarly as the original study \citep{2018carniania}. Normal star-forming galaxies from the ALPINE survey \citep{2020schaerer} are represented as black circles. The black solid line and filled area describes the best-fit for the distribution of galaxies at z $\sim$ 0 \citep{2014delooze}. \textbf{(d) Bottom right panel:} $\Sigma_{[CII]}$-$\Sigma_{SFR}$ relation. The lines and points are the same from the upper right panel, except for the black solid line and filled area that corresponds to the local $\Sigma_{[CII]}$-$\Sigma_{SFR}$ relation by \citep{2015herrera}. The dashed black line represents the best-fit for the galaxies and its respective clumps \citep{2018carniania}.}
    \label{fig:L[CII]-SFR-sigma-sigma}
\end{figure*}

\section{Discussion}
\label{sec:discussion}

\subsection{IR emission}


We have not found a detection of FIR continuum (Appendix ~\ref{appendix:data-products}), leading to an upper limit (3$\sigma$) for the IR luminosity of L$_{IR}^{3-1100{\mu}m}$ $\leq$ 1.15 $\times$ 10$^{10}$ L$_{\odot}$, assuming a single-component modified black body (MBB) approach and considering $\rm \beta_{IR}$ = 1.6 and T$_{\rm dust}$ = 53 K \citep{1989eales,1997klaas,2020bouwens}. This infrared luminosity can be translated into SFR$_{IR} \lesssim$ 1.7 M$_{\odot}$ yr$^{-1}$, for the calibration provided by \citet{2012kennicutt} (Kroupa IMF), which is 9 times lower than that estimated for the UV emission (see Section~\ref{sec:sigma}). Despite the non-detection of the dust continuum emission, the galaxy shows a red UV continuum slope of $\beta_{UV}$ = -1.18 \citep{2018smit}, which indicates some dust absorption. This value is also much redder compared to $\beta_{UV} \sim$ -2, which is the mean value expected for galaxies at z $\sim$ 7 \citep{2013dunlop,2014bouwens,2016bouwens}. This suggests that COS2987 has either a slightly older stellar population than other galaxies at this redshift and/or is more dust obscured. 

The low infrared luminosity points to a lower IR excess (IRX=L$\rm_{IR}$/L$_{UV} \leq 0.12$) compared to the upper limit found by \citet{2018smit}. A similar non-detection of dust emission is seen in normal galaxies at z = 5-6 \citep{2015capak}, and it could indicate a warmer dust, which can underestimate the infrared luminosity based in a MBB approach.  It follows the conclusion of \citet{2020fudamoto} and previous studies, which indicate a fast evolution of the ISM conditions in the first 4 billion years of the Universe. An alternative scenario for the non-detected dust emission is that the galaxy has a dust cold enough to be completely diluted by the cosmic microwave background, $\rm T_{CMB}(z = 6.8076) \sim 22\,K$. However, we are limited to infer how cold the dust needs to make undetectable the continuum at 158$\rm{\mu}m$, because this is strongly depended on the optical depth assumption for the gas.  

Finally, the galaxy has a lower-limit line to continuum luminosity ratio of \cii/FIR $\geq$ 0.015, comparable to values found for nearby low-metallicity dwarfs and star-forming galaxies \citep[0.1\% - 1\%,][]{2018herrera-camus}. A similar finding was obtained for the non-dust-detected normal galaxies at z = 5 - 6 from \citet{2015capak} (mean $\rm log\cii/FIR \sim -1.75$ for the typical upper-limits of  $\rm L_{IR} \lesssim 10.3$). Such high ratios suggest a lower dust-to-gas ratio and/or more diffuse clouds than normal galaxies at $z \sim 0$.
Our results for COS2987 further confirms that the dust continuum emission is difficult to detect in galaxies at the epoch of Reionization, even with the new ALMA sensitive observations. We still need to test the hypothesis that low metallicity or warm/cold dust components are the main culprits for such lack of detections.

\subsection{Resolved $\Sigma_{SFR}$ vs $\Sigma_{CII}$}
\label{sec:sigma}
In the past decade, a tight correlation between the \cii luminosity and the total SFR in nearby, un-obscured star-forming galaxies has been established \citep[z $\sim$ 0,][]{2011delooze,2015herrera}. The efforts to provide a similar relation for the early Universe reveal that this relation still exists \citep[e.g.,][]{2020schaerer}, but it features a higher dispersion \citep{2018carniania}. This suggests that the physical conditions of high-redshift galaxies may vary significantly from galaxy-to-galaxy, affecting how we link the star forming activity to the ISM conditions. Furthermore, the observed relation of the SFR and \cii luminosity surface densities in high-redshift galaxies shows an offset from the local relation, indicating a possible \cii deficiency and/or a more extended \cii effective radius with respect to the star forming components. 

We performed a global (integrated) and resolved analysis of the relation between the SFR and \cii luminosity and surface densities in COS2987. We defined apertures throughout the source and measured the corresponding fluxes (luminosities)  in the ALMA \cii and HST/F125W rest-UV maps. Given the lack of dust continuum detection in our data, we assumed that the SFR is largely dominated by the rest-UV emission, and thus SFR$_{\rm tot} \simeq$ SFR$_{\rm UV}$. We used the flux measured in the HST/F125W band as a tracer of the rest-UV emission, and converted the corresponding luminosity to SFR using the calibration provided by \citet{2012kennicutt} and a Kroupa IMF \citep{2003kroupa}. For the global measurement, we defined an aperture of 0.5\arcsec  radius (2.7 kpc), which roughly corresponds to the size of the central region of the galaxy C (as defined in Section~\ref{sec:spatial-distribution}). For the resolved measurements, we divided the source in 9 subregions, defined by apertures of $0.2\arcsec$ radius. The central and subregions are labeled from 1 to 9 in Figure~\ref{fig:L[CII]-SFR-sigma-sigma} top and bottom left panels. The circle labeled as 9 corresponds to the foreground galaxy discussed in Section \ref{sec:spatial-distribution},  therefore we discard any further analysis for this region. 

Figure \ref{fig:L[CII]-SFR-sigma-sigma} (top and bottom right panels) shows the global (dashed-yellow) and subregion (solid-red) apertures along with the measured L$_{[CII]}$-SFR and $\Sigma_{[CII]}$-$\Sigma_{SFR}$ relations of these regions. For comparison, we displayed measurements  from \citet{2018carniania} (C18, in Fig. \ref{fig:L[CII]-SFR-sigma-sigma}) in normal star-forming galaxies at $z = 5-7$ and from \citet{2020schaerer} in ALPINE star-forming galaxies at $z=4-6$. In the former case, measurements are provided for the full single systems (represented by red circles in Fig. \ref{fig:L[CII]-SFR-sigma-sigma}) and their clumps (yellow stars), with the SFR defined as SFR$\rm _{UV}$ + SFR$\rm _{IR}$ (SFRs of galaxies with no dust continuum detection are defined as SFR $\sim$ SFR$\rm _{UV}$). For the ALPINE galaxies, the only integrated measurements of each galaxy are provided, and the SFRs were obtained by SED fitting.

Our target, COS2987, is shown as a purple diamond and the resolved measurements as thin diamonds. Since some apertures (labeled as 2, 4, 6 and 8; Figure \ref{fig:L[CII]-SFR-sigma-sigma} - left panels) are overlapped with the central one (labelled as 5) and may have considerable emission from the central region, we separate them in non-filled thin diamonds. The other spatially independent apertures (labelled as 1, 3, 5, 7 and 9; Figure \ref{fig:L[CII]-SFR-sigma-sigma}) are shown as green filled diamonds. The galaxy lies within the nearby \cii-SFR relation (Figure \ref{fig:L[CII]-SFR-sigma-sigma}; top right panel).  It also holds for the resolved measurements, which follow a similar trend of increasing \cii luminosity with increasing SFR. This subregions are concentrated at a particular \cii - SFR range, indicating a homogeneous, non-complex structure. This result for the single system is consistent to the previous finding from \citet{2018smit}.

The bottom right panel of Fig.~\ref{fig:L[CII]-SFR-sigma-sigma} demonstrates how the spatial distribution is correlated with the \cii luminosity and SFR. The positive correlation of galaxies at $z=5-6$ is displaced from the local relation, such that galaxies and their substructures have a lower \cii luminosity than previously expected \citep[dashed black line;][]{2018carniania}. The \cii and SFR surface densities of the integrated and resolved measurements of COS2987 are defined at the effective radius of the \cii and UV emission, respectively. In this case,  both the global system and subregions locate slightly below (within 3$\sigma$ of) the local $\rm\Sigma_{[CII]} - \Sigma_{UV}$ relation \citep[black solid line and gray area;][]{2015herrera}. These results for both relations support the scenario shown by previous studies, where a \cii deficiency may be in place in high redshift galaxies, even for galaxies which follow tightly the local L$_{\cii}$-SFR relation. In any case, COS2987 and its subregions are closer to the local relation than we would expect from the high-z relation from C18, hinting that the physical conditions are closer to local galaxies than previous detections. 

The high line-to-continuum ratio  \cii/L$_{\rm IR} \geq 1.5\%$ and the fact that COS2987 lies within the local L$_{\cii}$-SFR relation evidence different physical mechanisms causing the \cii deficit than those taking place in high-infrared-luminosity starbursts \citep[$\rm\cii/FIR \sim 0.1\% - 0.01\%$][]{2018herrera-camus}. The \cii deficiency could be a consequence of an underestimation of the L$_{\cii}$, as \citep{2020carniani} reported fluxes losses of 20-40\% when the angular resolution is comparable to the size of the emitting region. Our high-resolution observations were combined to the shorter baselines of the previous low-resolution data set, and the latter has a $0.8\arcsec$ resolution sufficiently greater than the typical \cii emission size of COS2987 ($\sim0.5\arcsec$ obtained by a gaussian fit in the moment-0 map). Therefore, we conclude that this observational feature has no influence on the \cii deficiency in COS2987.

Recent PDR models and simulations attribute the existence of \cii under-luminous normal galaxies as a consequence of multiple conditions in the ISM such as a positive deviation from the Kennicutt-Schimidt relation (due to a starburst phase), low metallicity and low gas density \citep{2019ferrara}. Unless $Z < 0.1 $ Z$_{\odot}$, the metallicity and low gas density just play a sub-dominant role in the \cii deficiency \citep{2015vallini,2018lagache, 2019pallottini,2020lupi}. Any combination of these effects could be at play in COS2987, although such low metallicity (Z < 0.1 Z$_{\odot}$) would move COS2987 below the local L$_{\cii}$-SFR relation. 

In any case, the \cii emission can not fully characterize the ionization state of COS2987 and thus disentangle the degenerate contribution of these parameters. Therefore, other FIR lines, such as $\oiii_{88{\mu}m}$ and \ciii, are necessary to entirely solve this tension \citep{2020vallini,2020carniani, 2021vallini}.

\subsection{Dynamical state and evolution of $\sigma$}
\label{discussion-dynamics}

\begin{figure*}[ht]
    \centering
    \includegraphics[width=\textwidth]{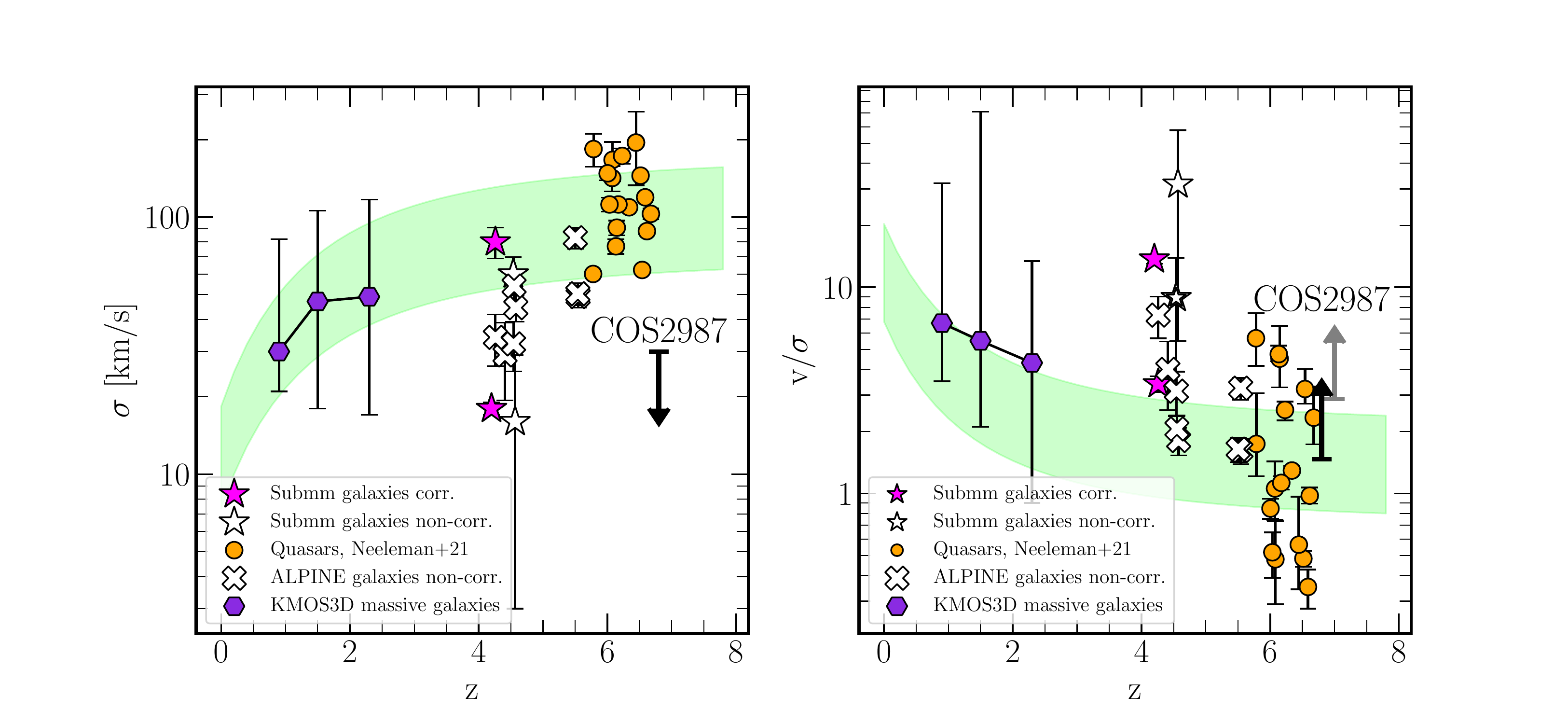}
    \caption{The dynamical state of COS2987 in the context of the cosmic evolution of the expected dispersion velocity (left panel) and rotation-to-dispersion ratio (right panel). As a reference, other intermediate- and high-redshift sources are displayed in both panels: KMOS3D massive galaxies at z = 1-2 \citep[purple hexagons;][]{2017ubler}, normal star-forming galaxies of ALPINE survey at z = 5-6 \citep[white crosses;][]{2021jones}, massive starbursts at z $\sim$ 4.5 \citep[pink and white stars;][]{2020neeleman,2021fraternali,2020rizzo} and quasars at z $>$ 6 \citep[orange circles;][]{2021neeleman}. The green filled-area represents the expected values for the dispersion and rotation-to-velocity from a semi-analytical model based on the Toomre parameter \citep{2015Wisnioski}. COS2987 is represented by upper-limits due to resolution limitation (see text). In the right panel, the black upper limit represents the rotation-to-dispersion ratio for the outer regions of the galaxy. The gray upper limit corresponds to the ratio for the average rotational and dispersion velocity. It is slightly offset from the black upper limit for better viewing.}
    \label{fig:dynamical-evolution}
\end{figure*}

\subsubsection{Alternative scenarios}

In Section ~\ref{sec:results}, we characterized the velocity map of COS2987 with a velocity gradient (Figure~\ref{fig:moment-1-2}) and a homogeneous velocity dispersion map typical of rotating-disk galaxies, as previously pointed out by \citet{2018smit}.  Nevertheless, this might not exclusively be produced by a rotating pattern, especially when it is associated with limited-resolution data. We thus consider two other possible origins to describe the observed velocity gradient in the galaxy: (i) flux of gas towards the circumgalactic medium caused by stellar and AGN feedback; and (ii) major mergers. If not satisfactorily spatially resolved, both cases can reproduce in the velocity map blueshifted and redshifted lobes smoothed by the beam size.  We will discuss the reality of these scenarios:

(i) Outflows can increase the dispersion of the system and add a broad component to the spectrum. Resolved observations and stacking spectrum of normal star-forming galaxies at z $>$ 5 show an extra component with a typical FWHM of few hundreds of  km s$^{-1}$ \citep{2020ginolfi,2021herrera-camus}. To test this, we fit a two-component Gaussian profile (broad and narrow component) to the spectrum of the main source C (top panel, Figure~\ref{fig:UV-[CII]-spectra}), but the resulting chi-square is indistinguishable from the fitting of a single Gaussian profile fit. Given the overall narrow line width of the [CII] line and the lack of evidence for a two component line profile, we consider this possibility unlikely. 

(ii) The large gas motions resulting from galaxy mergers can imprint multiple components on the moment-0 map and the spectrum, and  disturb  considerably the velocity dispersion map \citep{2021romano}. In the case of COS2987, we do not see any obvious secondary source component distribution in the \cii moment-0 map, spectrum and individual channel maps. Similarly, there is no evidence for merger activity in the central region of this source (C) in any of the available HST images. The latter is clearly described by a single Sersic profile (Fig.~\ref{fig:SB-profile}). Given the lack of dust detection at this position, it is unlikely that the rest-UV distribution seen in the HST images are affected by differential dust obscuration. The dispersion map shows a mainly homogenous low average dispersion $\sim$ 29 km s$^{-1}$, compared to merging systems at z $>$ 5 in the ALPINE survey \citep{2021romano}. Moreover, COS2987 has a lower mass of 1.7 $\times$ 10$^{9}$ M$_{\odot}$ compared to the mass of $\sim 10^{10}$ M$_{\odot}$ of these merging systems. With the current ALMA [CII] and HST rest-UV data, we are able to discard a compact merger with separations larger that 2-3 kpc.

As suggested by \citet{2022rizzo} after several tests with disks and mergers simulations, an accurate differentiation of disks and mergers galaxies requires observations to cover three beams over the major axis of the galaxy and S/N $>$ 10, conditions which are not reached by our data. With this caveat in mind, in the next two subsections we examine the implications of an evolution of disk galaxies assuming that the velocity gradient in the velocity map and p-v diagram arrive from a rotating gaseous disk.

\subsubsection{Stability and evolution}

The evolution of the intrinsic velocity dispersion and the rotation-to-dispersion ratio with redshift elucidate which physical mechanisms can play a role during galaxy evolution \citep[][ and references therein]{2018girard, 2019ubler}. Both simulations and observations point to an increase of the velocity dispersion with redshift indicating that high-redshift disks galaxies are ``hotter'' dynamically \citep{2019Pillepich}. Galaxies may have experienced constant turbulence caused by several physical mechanisms, such as: energy transfer by winds due to supernova episodes that can generate gas outflows to the outer regions \citep{2004maclow}, instability due to the accretion process of cold gas from cosmic filaments \citep{2009dekel} and instabilities generated by major and minor mergers \citep{2010Bournaud,2019Pillepich}. Furthermore, recent simulations indicate that rotation-supported systems should be expected in the first billion years after the Big Bang\citep{2005kerev,2009dekel}. 

Sensitive CO and H$\alpha$ imaging surveys of main-sequence galaxies at $z\sim1-2$ have shown that this population is dominated by rotation-supported systems \citep{2010tacconi, 2019ubler, 2020genzel, 2021sharma}. Recent studies, mainly based on \cii analysis, point to the presence of fast cold rotators among massive galaxies with intense star formation already at z $\sim$ 4.5 \citep{2020neeleman,2020rizzo,2021fraternali,2021lelli}. Furthermore, analysis of ALPINE \cii data cubes  indicates that an important fraction ($\sim40\%$) of the massive main sequence galaxies at $z=4-6$ can also be associated to rotating disks \citep{2021jones} (z = 4 - 6).  \citet{2018smit} showed that rotating disks are already in place at $z\sim7$. One of the big challenges is the limited resolution and depth of these studies to confirm whether these systems are in fact rotating disks.

If we consider the velocity gradient of COS2987 comes from a rotating-disk, we can check the stability state of the system, and put it into context of the dynamical evolution of disks. In Fig.~\ref{fig:dynamical-evolution}, we show the intrinsic velocity dispersion and rotation-to-dispersion ratio as a function of redshift for COS2987 and recent observations. We compared them with the expected average dynamical parameters for galaxies from the semi-analytical model of \citet{2015Wisnioski, 2019wisnioski} (green area). We included a sample of main-sequence star-forming galaxies from the ALPINE survey \citep[white crosses,][]{2021jones} at $z>4$ and the following massive starburst galaxies:  DLA0817g  \citep[z = 4.26,][]{2020neeleman}, SPT–S J041839–4751.9  \citep[z = 4.2,][]{2020rizzo}, and finally AzTEC/C159 (z = 4.57) and J1000+0234 \citep[z = 4.54,][]{2021fraternali}. They are all signed as stars in Figure~\ref{fig:dynamical-evolution}.  We also included available dynamical data from quasar host galaxies at $z > 6$ \citep[orange circles;][]{2021neeleman}. 

We noted that there are differences in the definitions of the rotational and dispersion velocity in the literature. We homogenized them as specified by \citet{2015Wisnioski}, in which the velocity dispersion is defined at the  outer regions of the galaxy to avoid beam smearing effects. The quasars \citep{2021neeleman} and the massive galaxy DLA0817g \citep{2020neeleman} were derived by assuming a constant dispersion and rotational velocity profile, therefore no change is needed. For the starburst massive galaxy SPT0418-47, \citet{2020rizzo} assumed an exponential model for the dispersion velocity profile. We thus used the measurements for R $>$ 1 kpc, where the rotational velocity and dispersion flatten. These two massive galaxies are represented as pink colors in Figure \ref{fig:dynamical-evolution}. For the massive galaxies  AzTEC/C159 and J1000+0234 \citep{2021fraternali} and the normal star-forming ALPINE galaxies \citep{2021jones}, the rotational and dispersion velocities were obtained using \barolo and no fixed radial profile shape for the velocity dispersion was used. Due to the beam size, the points for the rotational and dispersion velocities profiles are not independent. For each galaxy, we thus simply took the average value in all rings and caution of possible conclusions for these systems. Thefore, all cases that could not be corrected are marked in white.  Finally, for the galaxy COS2987, we computed values for 3 rings with a width of 0.18\arcsec, and we took the value for the third ring, located at 0.45\arcsec (black upper limit in both panels of Fig. \ref{fig:dynamical-evolution}). However, since it is close to the \cii effective radius (0.38\arcsec), and the points are not completely independent, we also display in the left panel the average rotational-to-dispersion ratio (gray upper limit slightly offset from the black one for better visualization).

From this comparison, we note that the model agrees with the average dynamical state of massive disks at $z\sim1-2$ \citep[KMOS3D survey;][]{2017ubler} and quasar host galaxies at z $\sim$ 6. However, it struggles to reproduce the velocity dispersion and rotation-to-dispersion ratios of most massive and star-forming galaxies at $z\sim4-6$. The typical star-forming galaxies and some of the massive galaxies may be affected by beam smearing, which can overestimate the values of the rotation-to-dispersion ratio. But the massive starburst galaxies, specially SPT0418-47 ($\sigma = 18 \pm 1$ km s$^{-1}$, and $ \rm v/\sigma = 13.7 \pm 0.7$), also diverge from what is expected by the models and simulations.

We found a value of v$_{rot}$/$\sigma \geq$  1.4 for COS2987, which agrees with the semi-analytical models, but we noted a low velocity dispersion (upper limit, Fig \ref{fig:dynamical-evolution} - right panel),  where the expectation is to have highly turbulent systems. This implies that the disk, if in place, is likely in a stable dynamical state. This is mildly at odds with our finding of a northern arm, which could represent an on-going minor merger (and/or gas inflow), and thus imply a highly active stage of assembly. Such an event (and possibly latter ones), would result in a significant perturbation of the galaxy disk. As pointed out by \citet{2019katz}, observing stable disks at these epochs might represent a short-lived phase, likely coincidental with a particular line-of-sight and geometry, before the galaxy experiences further perturbations. In the case of quasars, it is likely that such objects are in a highly turbulent phase, due to the presence of a dominant AGN.

We noted that the choice of  source inclinations for COS2987 would increase the resulting velocity dispersion, thus providing a better agreement with the expected dispersion velocity at z $\sim$ 7.  To test this, we ran \barolo with a fixed dispersion velocity of 65 km s$^{-1}$, which matches the expected velocity dispersion at this redshift. Higher inclinations yield poorer fits to the velocity dispersion maps, making such conclusion unlikely. These results rule out the possibility that our galaxy would represent a highly unstable system. 

We diagnosed the prospect of the candidate gaseous disk to develop local instabilities by the Toomre Q-parameter \citep{1964toomre}. It is defined by $Q = \kappa \,\sigma/(\pi  \, G \, \Sigma_{gas})$, where $\sigma$ is the dispersion velocity; G is the gravitational constant; $\Sigma_{gas}$\footnote{$\Sigma_{gas}$ = $\frac{M_{gas}}{\pi (r^{eff}_{[CII]})^{2}}$} is the gas surface density and $\kappa = (2\Omega/r)d(r^{2}\Omega)/dr = av_{c}/r$ is the epicyclic frequency, where $v_{c}$ is the circular velocity and $a = \sqrt{2}$ for a flat rotational curve. For the gas surface brightness, we adopted the total gas mass at the effective radius of M$_{gas}$ = (5 $\pm$ 3) x 10$^{9}$ M$_{\odot}$ (M$_{gas} \sim$ M$_{dyn}$  - M$_{*}$/2, M$_{*}$ see Subsection ~\ref{sec:conversion-factor}). If Q < 1, the gaseous disk is susceptible to local gravitational collapse. Otherwise (Q $\geq$ 1), the system would not be inclined to develop clumpiness. For the effective radius and maximum rotational velocity, we found a mean Q = 0.44, indicating that even thought COS2987 is a possible rotationally-supported system, it is prone to face gravitational instabilities in its internal structure.

\subsubsection{Conversion factor}
\label{sec:conversion-factor}
We dynamically characterized the galaxy for a rotating-disk scenario, therefore it is possible to measure the dynamical mass of the system within its effective radius. Rest-frame near-infrared observations provides the measurement of the extension of the stellar population of the galaxy. Since the rest-frame optical data from Spitzer imaging does not resolve the galaxy sufficiently, we chose the \cii effective radius as a proxy of the stellar extension, supported by the agreement of the \cii and UV effective radii. From the maximum rotation velocity of the system, we obtain a dynamical mass of M$_{dyn}$ =(6 $\pm$ 3) x 10$^{9}$ M$_{\odot}$. We estimate the total mass of gas at the effective radius given that the inner region of the galaxy is baryonic dominated \citep[M$_{gas} \sim$ M$_{dyn}$  - M$_{*}$/2, M$_{*}$ = $1.7$ x $10^{9}$ M$_{\odot}$;][]{2018smit}, leading to the value of M$_{gas}$ = (5 $\pm$ 3) x 10$^{9}$ M$_{\odot}$.

Considering the \cii  luminosity within the \cii effective radius of the galaxy (0.38\arcsec $\sim$ 2 kpc), we get the \cii-to-gas conversion factor. We estimate a conversion factor for the total galaxy gas mass as $\alpha^{total}_{\cii}$ = 62 $\pm$ 7 M$_{\odot}$/L$_{\odot}$ which is a considerably larger value than the median (7$^{
+4}_{-1}$ M$_{\odot}$/L$_{\odot}$) for  starburst galaxies supported by rotation at z = 4.5 \citep{2021Rizzo}.

Finding the conversion factor for molecular gas is a harder task, since there is no direct measurement of the molecular gas mass at this redshift. If we assume that 70\% of the \cii emission is due to PDR regions \citep{1991stacey,2010stacey}, we find an upper limit of  $\alpha^{mol}_{[CII]} \leq$ 90 M$_{\odot}$/L$_{\odot}$\footnote{Derived from $\alpha^{mol}_{[CII]} = \frac{M_{gas}}{(70\% \times L_{[CII]}/2)}$ .}, a higher value than the one achieved by \citet{2018zanella} (mean value of $\alpha^{mol}_{[CII]} = 31$ M$_{\odot}$/L$_{\odot}$).


\section{Conclusions}
\label{sec:conclusions}
In this paper, we analyze the morphology and kinematics state of the normal star-forming galaxy COS2987, at z $\sim$ 7. Previous studies found a low dust content and a tentative evidence for a velocity gradient suggesting possible disk rotation (but indistinguishable from a merger scenario). The new higher-resolution \cii observations obtained with ALMA in Band 6 can resolve 2 kpc-scale structures, allowing us to perform a detailed study of the cold ionized gas conditions, distribution and kinematics. We compared the spatial distribution of the stellar and \cii line emission components to search for hints of extended \cii emission. We also performed a dynamical fitting using \barolo to study the possibility of rotation in the galaxy and put in context with the cosmic evolution of disk galaxies. Finally, the high resolution allowed us to look individually at the components of the galaxy and study the scattering of the L$_{CII}$-SFR relation and offset in the $\Sigma_{SFR}$ vs $\Sigma_{CII}$ relation.

We summarized our findings below:

\begin{itemize}
    \item \textbf{No extended emission/\cii halo:} COS2987 presents a good match between the cold ionized gas (\cii) and UV continuum distribution. The \cii and UV spatial extension agrees within the uncertainties and the radial profile of the emissions is similar in shape and size, even thought we could see a \cii emitter satellite in the north-east portion of the galaxy. Therefore, we did not detect signs of a \cii halo, indicating that such structures are not ubiquitous at these redshifts. 
    
    \item \textbf{Complex environment:} The galaxy presents an extended arm-shaped structure, which can be associated to a satellite being accreted. Additional possible evidence for \cii structures and clumps are found in its neighborhood, within a 10-kpc region. Deeper observations are needed to confirm these structures. 
    
    \item \textbf{A disk candidate in the early Universe}: The rotation-to-dispersion ratio agrees with the  models, but a low velocity dispersion reveals that COS2987 is going through a non-turbulent phase.  If the scenario of having a complex environment composed of other satellites is correct, the stability of the candidate disk can be a short event, with the disk being affected in posterior times. The suggested rotating disk nature hints that these structures appear as early as the first billion year of the universe. But the low fraction of rotation ordered normal star-forming galaxies \citep[40\%;][]{2021jones} at $z > 4$ indicate that this is a subdominant dynamic state in this cosmic epoch. Deeper and higher-resolution observations are mandatory to rule out the scenario in which COS2987 is a compact merger. 
    \item \textbf{\cii deficiency:} COS2987 and its substructures lie along the tight correlation L$_{[CII]}$ and SFR of nearby normal star-forming galaxies, but they fall below the local relation of $\Sigma_{SFR}$ vs $\Sigma_{CII}$.  This suggests that the \cii emission may be affected by a strong radiation field caused by a starburst phase, low metallicity or low gas density and complementary FIR lines are needed to fully understand the ionization structure of COS2987.

\end{itemize}

Our findings support the necessity of kpc-scale resolved observations to unveil the nature and properties of the ISM in galaxies early in the Universe. Forthcoming similar observations in larger samples (e.g. like the REBELS and CRISTAL ALMA large programs) will provide insights on the ubiquity of disks at early cosmic times. Future observations with JWST will also help us to constrain the high-ionization optical lines, which are critical tracers of warm, moderate density gas environments.

\begin{acknowledgements}
A. P. and M.A. acknowledge support from FONDECYT grant 1211951, ``ANID+PCI+INSTITUTO MAX PLANCK DE ASTRONOMIA MPG 190030'', ``ANID+PCI+REDES 190194''. M.A. acknowledges partial support from ANID BASAL project FB210003. G.C.J. acknowledges ERC Advanced Grants 695671 ``QUENCH’’ and 789056 ``FirstGalaxies’’, as well as support by the Science and Technology Facilities Council (STFC). RJA was supported by FONDECYT grant number 1191124 and by ANID BASAL project FB210003. CR acknowledges support from the Fondecyt Iniciacion grant 11190831 and ANID BASAL project FB210003. RS acknowledges support from an STFC Ernest Rutherford Fellowship (ST/S004831/1). R.H.-C. thanks the Max Planck Society for support under the Partner Group project ``The Baryon Cycle in Galaxies" between the Max Planck for Extraterrestrial Physics and the Universidad de Concepción. R.H-C. also gratefully acknowledge financial support from Millenium Nucleus NCN19058 (TITANs), and ANID BASAL projects ACE210002 and FB210003. This paper makes use of the following ALMA data: ADS/JAO.ALMA\#2015.1.01111.S and ADS/JAO.ALMA\#2018.1.01359.S. ALMA is a partnership of ESO (representing its member states), NSF (USA) and NINS (Japan), together with NRC (Canada), MOST and ASIAA (Taiwan), and KASI (Republic of Korea), in cooperation with the Republic of Chile. The Joint ALMA Observatory is operated by ESO, AUI/NRAO and NAOJ. 
\end{acknowledgements}

\bibliography{bibli.bib}
\bibliographystyle{aa}

\appendix

\section{Data products}
\label{appendix:data-products}
\subsection{\cii moment-0 maps for the available observations}

\begin{figure*}[bp!]
    \centering
    \includegraphics[scale=0.34]{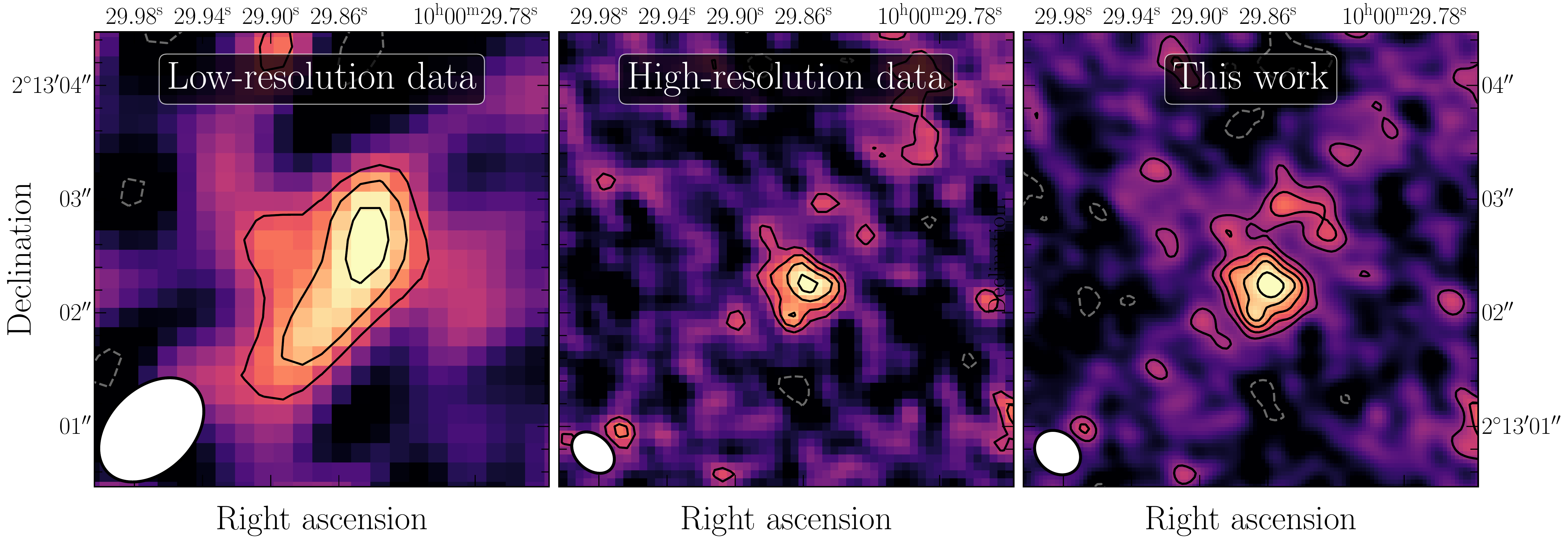}
    \caption{Comparison of the $5\arcsec\times5\arcsec$ zoom-in \cii moment-0 maps for the available observations of COS2987. The left, central and right panel correspond to the observation of \citet{2018smit}, our observation and the combination of both, respectively. The overlaid black contours show the 2, 4, 6 and 8-$\sigma$ levels, and the dashed dark gray contours show the -2$\sigma$ level of each \cii  map. The white ellipse in the bottom right corner represents the beam size.}
    \label{fig:comparison-old-new-data}
\end{figure*}

In Figure~\ref{fig:comparison-old-new-data} we display the \cii moment-0 maps for the available observations: low-resolution data from \citet{2018smit} (left panel; PI: Smit, ALMA ID: 2015.1.01111.S), high-resolution data obtained in Cycle 6 (central panel; PI: Aravena, PID: 2018.1.01359.S),  and the combination of both data sets (right panel). The cubes were obtained using the \textit{tclean} task of CASA software using natural weighting to preserve sensitivity. The maps represent the collapsed line cube, averaged over a 150 km s$^{-1}$ centered at the \cii line frequency.

\subsection{Data cube}

\begin{figure*}[bp!]
    \centering
    \includegraphics[scale=0.35]{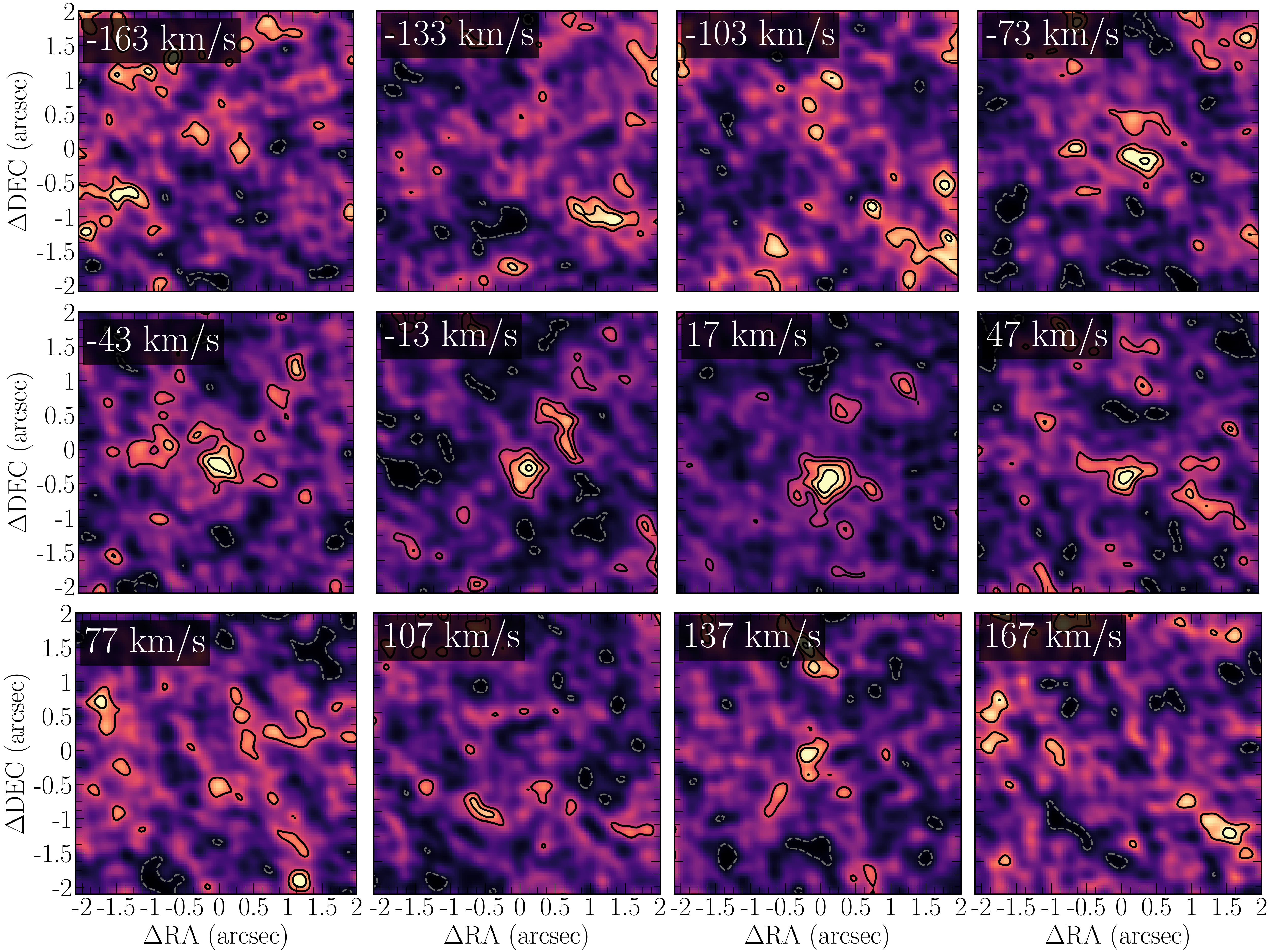}
    \caption{Channel maps of the cube within a velocity range of [-180,180] km s$^{-1}$  (in bins of 30 km s$^{-1}$), centered at the observed \cii frequency. The overlaid black contours show the 2-, 4-, 6- and 8-$\sigma$ levels of the maps, and the dashed dark gray contours show the -2$\sigma$ level of the \cii maps. The central velocity of each channel is placed in the left corner of each image. }
    \label{fig:central-channels}
\end{figure*}

In Figure~\ref{fig:central-channels} we provide the $2\arcsec\times 2\arcsec$ \cii channel map postage stamps in velocity steps of 30 km  s$^{-1}$, within the velocity range of [-180,180] km s$^{-1}$.

\subsection{Galaxy wide-field}

\begin{figure*}[bp!]
    \centering
    \includegraphics[scale=0.36]{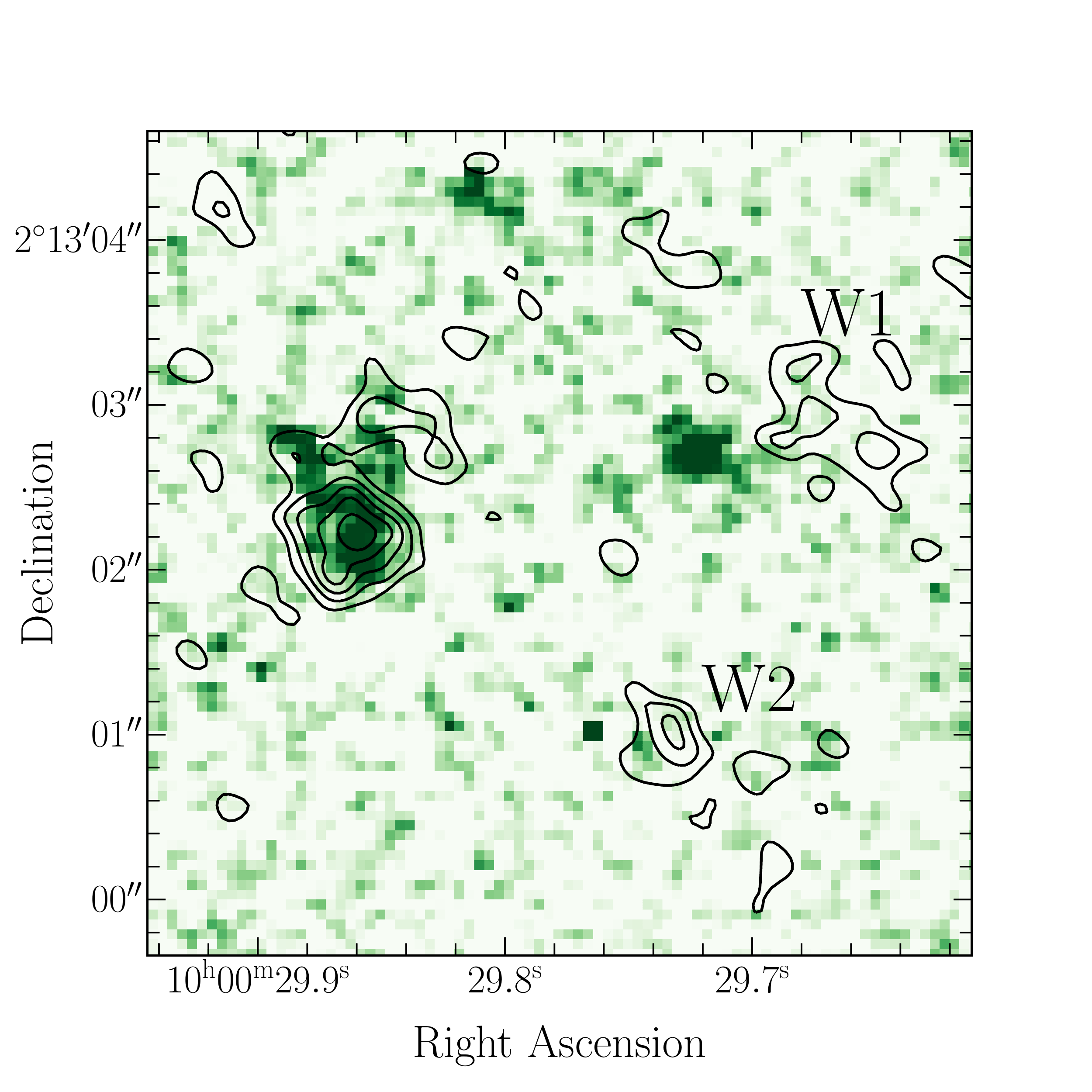}
    \includegraphics[scale=0.36]{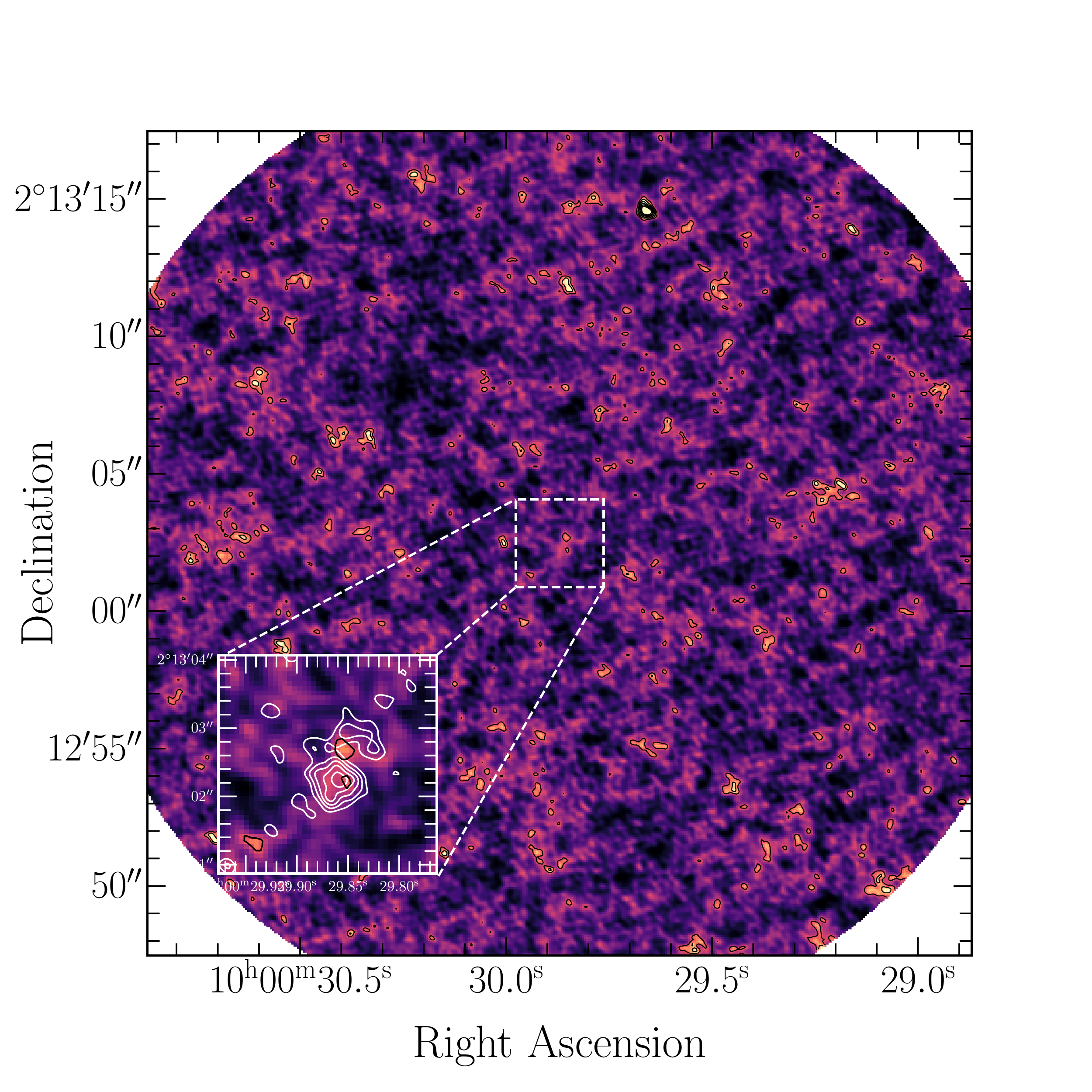}
    \caption{Wide-field rest-frame UV and continuum maps of COS2987 and surrounding regions. \textbf{(a) Left panel:} HST F160W-band image in a $6\arcsec \times 6\arcsec$ region containing the central source and the candidate \cii emitters (see Section \ref{sec:spatial-distribution}). The \cii moment-0 map is overlaid in black contours represented by the significance levels of 2-, 3-, 4-, 5-, 6-$\sigma$. \textbf{(b) Right panel:} Continuum map of the COS2987 field. The black lines correspond to the continuum significance levels from 2- to 7-$\sigma$ at steps of 1-$\sigma$. In the bottom left panel within the main panel, we zoom in the central source in a $4\arcsec\times4\arcsec$ region. The overlaid white lines represent the \cii moment-0 map for the significance levels from 2- to 6-$\sigma$, at steps of 1-$\sigma$. }
    \label{fig:appendix-f160w-dust}
\end{figure*}

In Figures \ref{fig:appendix-f160w-dust} left panel, we display the F160W-band image overlapped by the \cii moment-0 map in contours at significance levels of 2-, 3-, 4-, 5-, 6-$\sigma$. In Section \ref{sec:spatial-distribution} we searched for a UV counterpart for the candidate \cii emitter W1 and W2, and we just found a rest-frame UV emitter which is located east from W1. The centroid distance of 0.95\arcsec for the central source C and W1 translates into a projected physical separation of 5 kpc. It was previously identified as a galaxy with photometric redshift of 1.73 by \citet{2016laigle}. If the W2 candidate emitter was located at this redshift, the tentative emission would be explained by some emission at rest-frame frequency of 664.35 GHz. The closest bright possible emission would be the CO(6-5) at 691.47 GHz, therefore we ruled out this possibility.

In Figure \ref{fig:appendix-f160w-dust} right panel, we display the wide-field continuum image centered at the COS2987 galaxy. The image excludes any channel within the velocity range [-250,250] km s$^{-1}$ centered at the line. We zoom in the region represented by the white dashed line in the bottom left panel. The black solid lines correspond to the continuum significance levels from 2- to 7-$\sigma$ at steps of 1-$\sigma$ and the white solid lines correspond to the \cii emission significance level of 2-, 3-, 4-, 5-, 6-$\sigma$. We reported the continuum detection for the galaxy COSMOS 4104 ($\alpha$, $\delta$) =(10:00:29.6668, +02:13:14.591)  at the photometric redshift 2.25 \citep{2017nayyeri}.

\section{Sources properties}
\label{sec:appendix-sourceproperties}
In Section~\ref{sec:spatial-distribution} we described the properties of four \cii emitters: the central source (C), north-east emission (N1), north-west arm  (N2), west emission (W1),  south-west emission (W2). The latter two are deemed as candidate \cii emitters. They are shown in Figure~\ref{fig:UV-[CII]} (central panel) and their properties are listed in Table \ref{tab:properties-clumps}. The properties are taken from a region that contains the emission of each source. The central coordinates (right ascension and declination) represent the centroids of each emission. The \cii luminosity is obtained following \citep{1992solomon}, using:

\begin{equation}
    L_{[CII]} = 1.04 \times 10^{-3} \; S_{[CII]}  \Delta v \: D^{2}_{L} \: \nu_{obs},
\end{equation} 

where L$_{[CII]}$ is measured in L$_{\odot}$, the velocity integrated flux, S$_{[CII]} \Delta v$, in Jy km s$^{-1}$, the observed frequency, $\nu_{obs}$, in GHz, and the luminosity distance, D$_{L}$, in Mpc.

\begin{table*}[bp!]
\centering          
\begin{tabular}{l c c c c}     
\hline\hline       
Property &                                     N1              &  N2             & W1               & W2 \\ 
\hline                    
   RA                                       & 10:00:29.8841    &  10:00:29.8429  & 10:00:29.6692   & 10:00:29.7354\\  
   DEC                                      & +02:13:02.671    &  +02:13:02.871  & +2:13:02.832    & +02:13:00.958\\ 
   S/N                                      & 2.2 $\sigma$     &  4.1$\sigma$    & 5.2$\sigma$     & 3.6$\sigma$ \\ 
   \cii line flux (Jy km s$^{-1}$)          & 0.02$\pm$ 0.01   & 0.07 $\pm$ 0.02 & 0.10 $\pm$ 0.03 & 0.06 $\pm$ 0.02  \\ 
   L$_{[CII]}$ ($10^{8}$ L$_{\odot}$)       & 0.2 $\pm$ 0.1    & 0.8 $\pm$ 0.2   & 1.1 $\pm$ 0.3   & 0.7 $\pm$ 0.2 
   \\ 
 

\hline  
\hline
\end{tabular}
\caption{\cii properties of the identified clump candidates around COS2987. The location of the emitters are indicated in Figure~\ref{fig:UV-[CII]} (central panel).}
\label{tab:properties-clumps}
\end{table*}

\section{Target extension}
\label{sec:appendixsersic}
\begin{figure*}[bp!]
    \centering
    \includegraphics[scale=0.45]{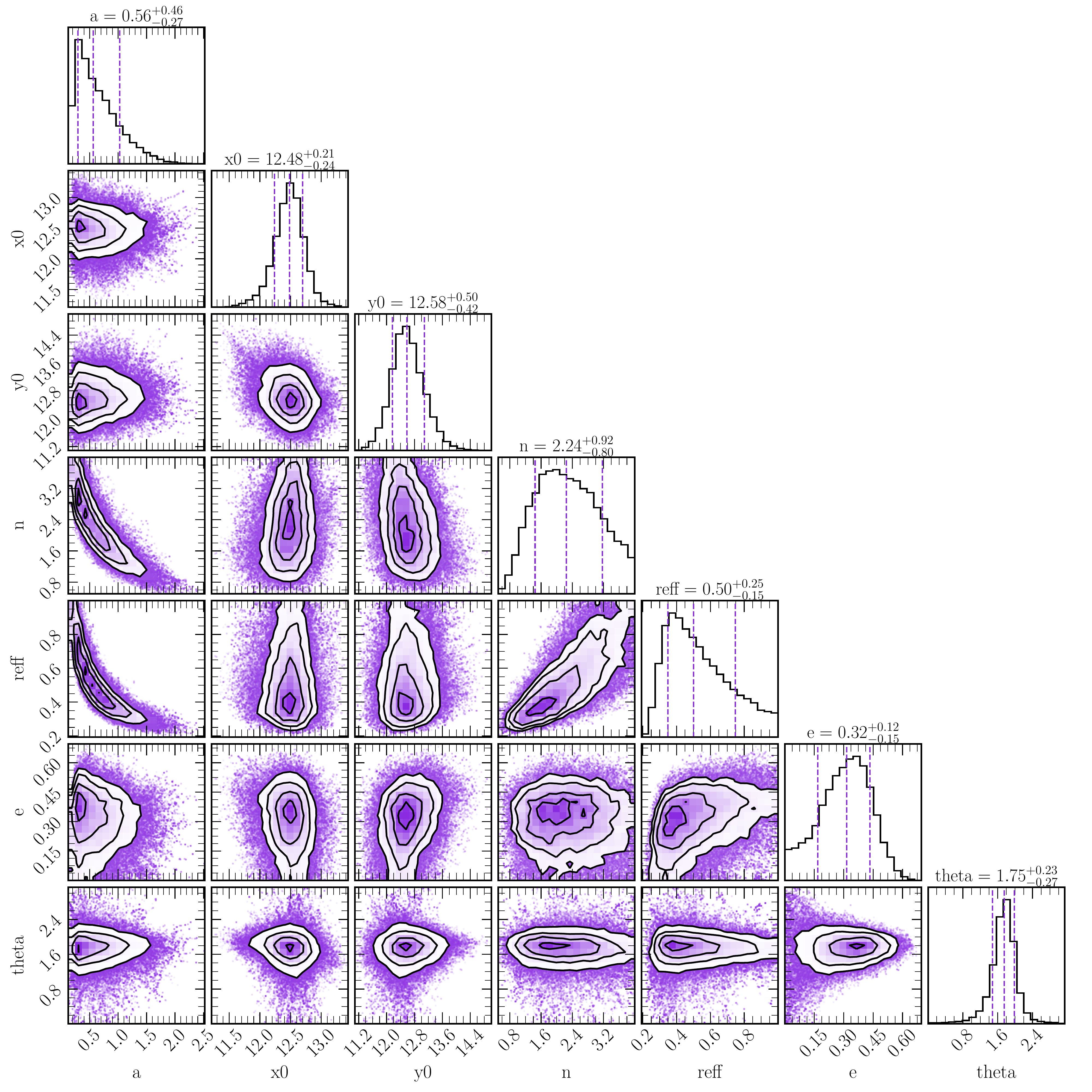}
    \caption{ Posterior distributions of parameters of the two-dimensional UV surface brightness, setting the S\'ersic index as a free parameter. The S\'ersic profile model also includes the surface brightness at the effective radius (a), the central pixel position ($x_0$ and $y_0$), the effective half-light radius ($r_{eff}$), ellipticity ($e$) and the rotation angle (theta). Black contours
correspond to 16\%, 50\% and 84\% confidence regions, shown as the dashed purple lines. }
    \label{fig:MCMC-HST-free}
\end{figure*}

\begin{figure*}[bp!]
    \begin{center}
    \includegraphics[scale=0.45]{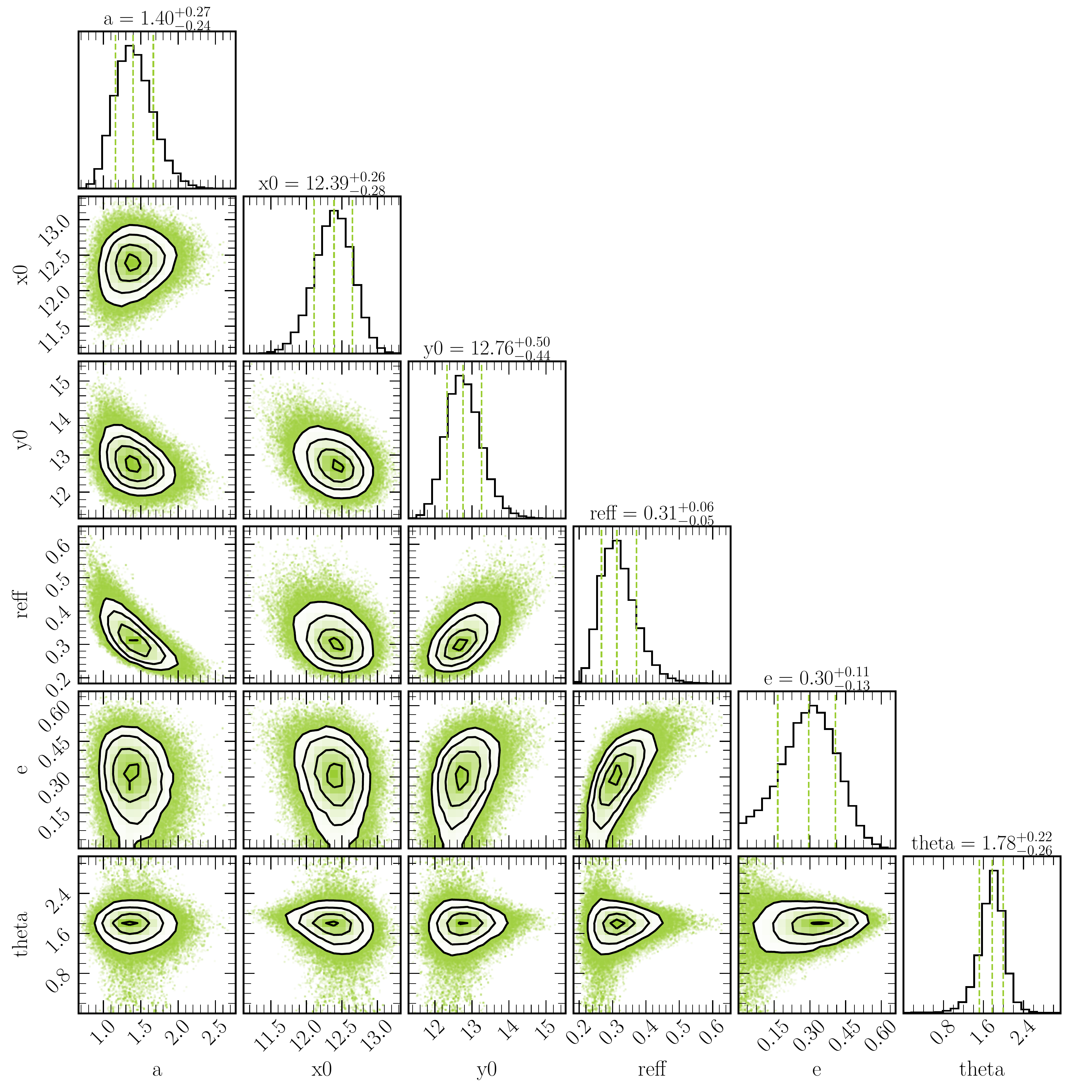}
    \caption{Same as Fig. \ref{fig:MCMC-HST-free}, setting the S\'ersic index equals 1. Black contours correspond to 16\%, 50\% and 84\% confidence regions, shown as the dashed green lines. }
    \end{center}
    \label{fig:MCMC-HST-1}
\end{figure*}

\begin{figure*}[bp!]
    \centering
    \includegraphics[scale=0.48]{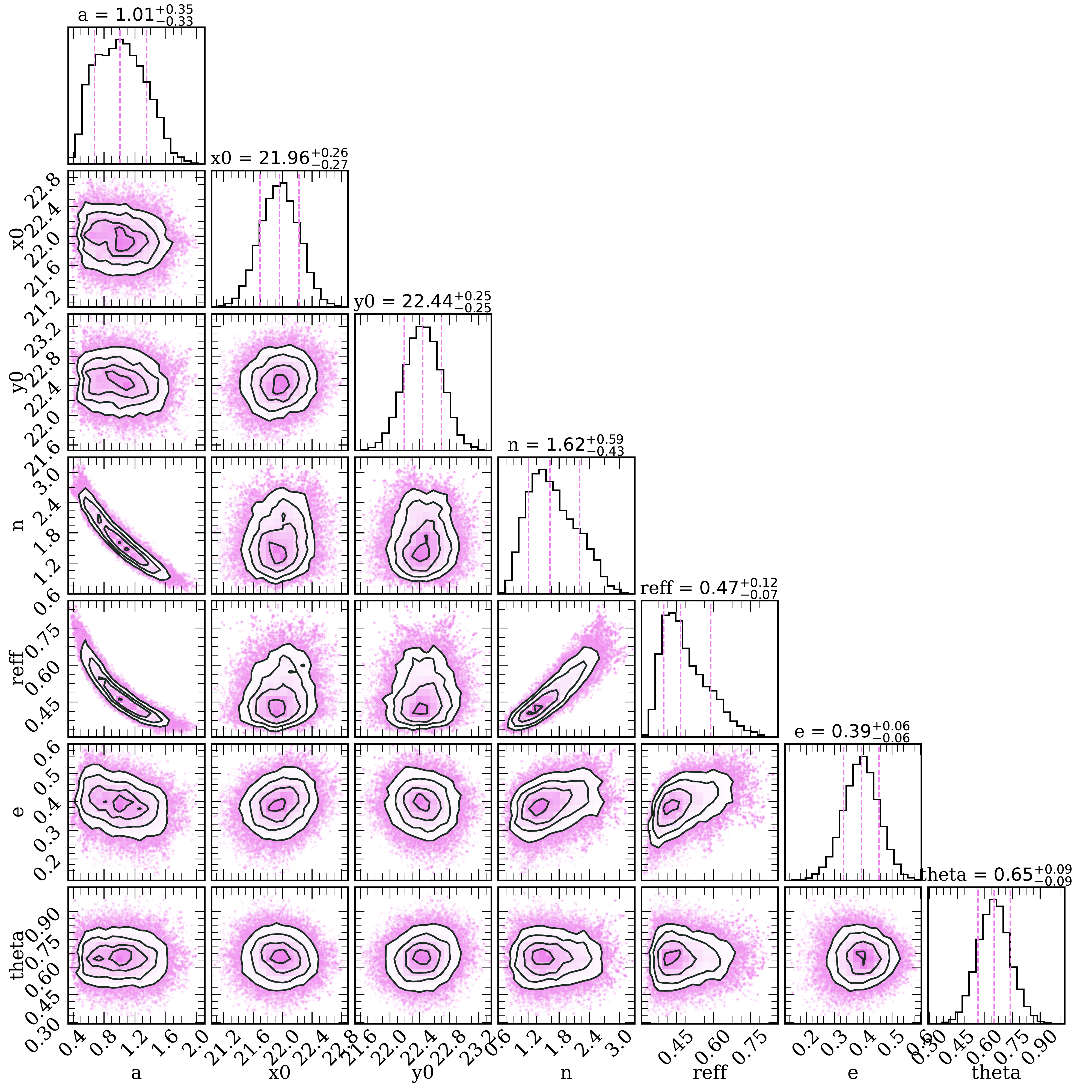}
    \caption{Posterior distributions of parameters of the two-dimensional \cii surface brightness, setting the S\'ersic index as a free parameter. The Sérsic profile model also includes the surface brightness at the effective radius (a), the central pixel position ($x_0$ and $y_0$), the effective half-light radius ($r_{\rm eff}$), ellipticity ($e$) and the rotation angle (theta). Black contours
correspond to 16\%, 50\% and 84\% confidence regions, so as the dashed pink lines.}
    \label{fig:MCMC-ALMA-free}
\end{figure*}

\begin{figure*}[bp!]
    \centering
    \includegraphics[scale=0.5]{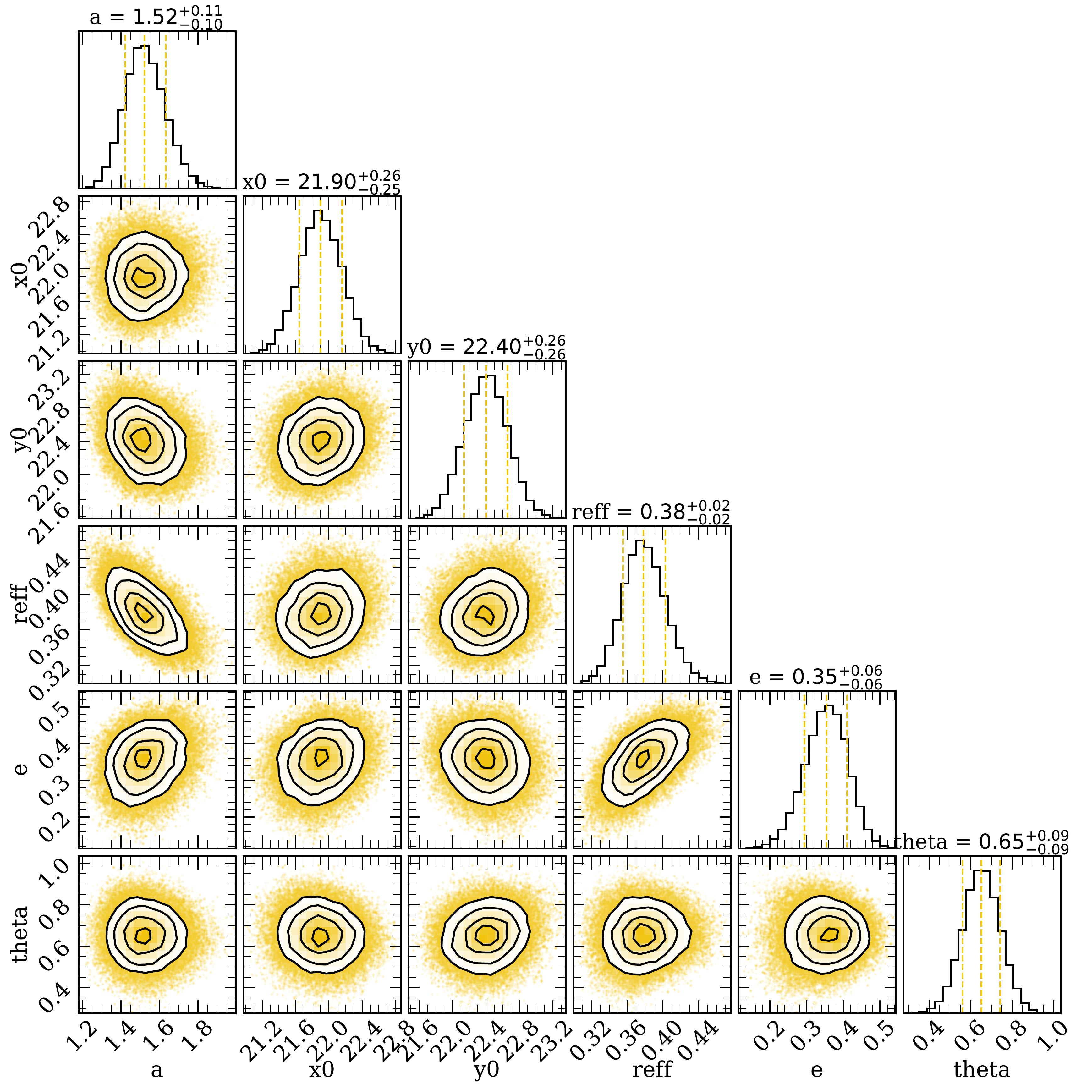}
    \caption{Same as Fig. \ref{fig:MCMC-ALMA-free}, setting the S\'ersic index equals to 1. Black contours correspond to 16 per cent, 50 per cent and 84 per cent confidence regions, so as the dashed yellow lines.}
    \label{fig:MCMC-ALMA-1}
\end{figure*}

We employed a Monte Carlo Markov Chain (MCMC) technique to fit a two-dimensional S\'ersic profile to the \cii and UV surface brightness distribution. We use the \textit{Sersic2D} task from the \textit{astropy} package in python to generate the surface brightness map, which is described by the following parameters: amplitude (\textit{a}; surface brightness at the effective radius), ellipticity (\textit{e}), S\'ersic index (\textit{n}), effective half-light radius ($r_{\rm eff}$), rotation angle ($\theta$) and central pixel position ($x_0$ and $y_0$). For the rest-frame UV and \cii emission, we follow two approaches: (i) setting all the parameters as free; and (ii) fixing the S\'ersic index $n = 1$, for a fair comparison with previous studies\citep{2020fujimoto}.  We excluded the pixels within the region corresponding to the north-east foreground galaxy \citep[located a z = 2.099;][]{2017laporte} and the north-west arm. In the case of ALMA maps, we assume a constant Gaussian noise across the whole image. In the case of HST images, we consider a background Gaussian noise and Poisson noise from the emission added in quadrature.  We ran the MCMC routine with 100 walkers for 5000 interactions. In Figures~\ref{fig:MCMC-HST-free}, \ref{fig:MCMC-HST-1}, \ref{fig:MCMC-ALMA-free} and \ref{fig:MCMC-ALMA-1} we show the posterior distributions of each parameter.

\section{\barolo products}
\label{appendix:trial-and-error}

\begin{figure*}[bp!]
    \centering
    \includegraphics[scale=0.55]{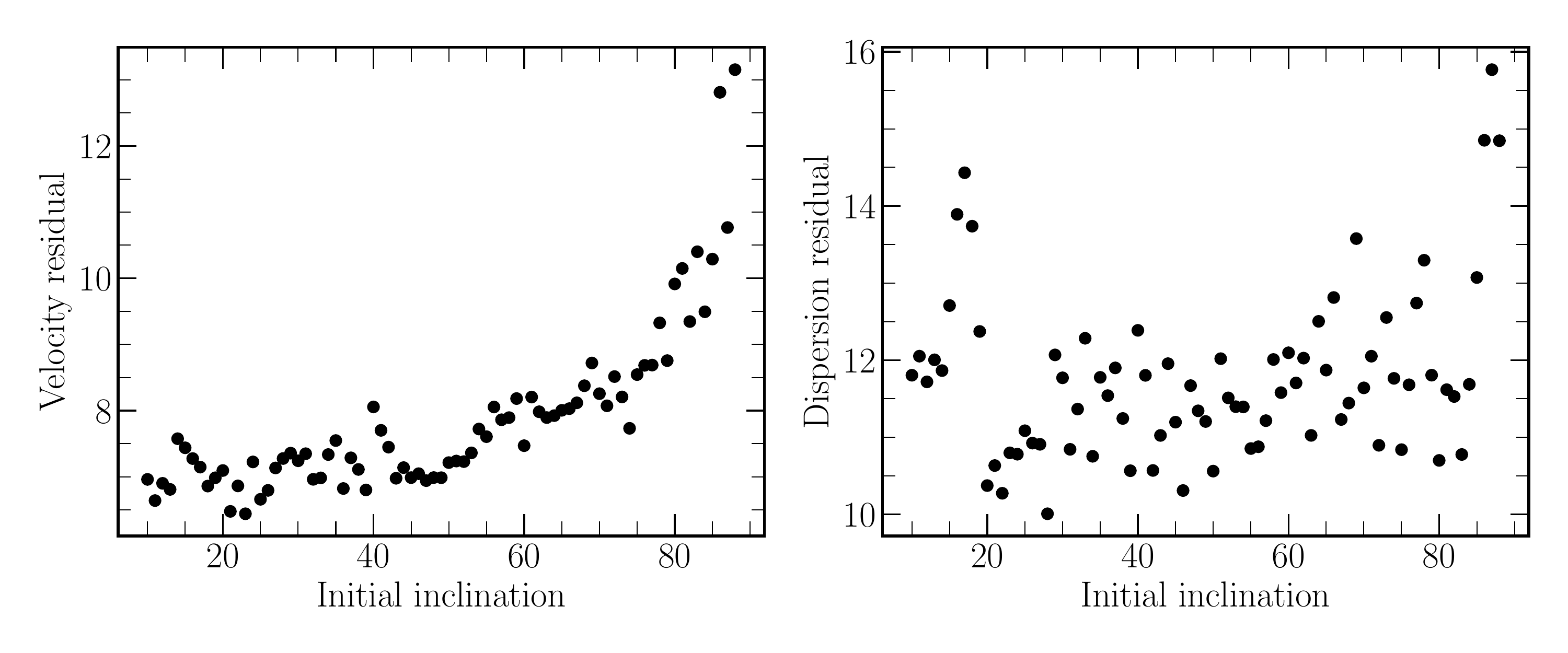}
    \caption{Residuals (Data - model, weighted by the \cii moment-0 map) of the trial-and-error tests for the radial (left panel) and dispersion velocities (right panel) maps for initial inclinations ranging from 10 to 90 $\deg$. }
    \label{fig:trial-and-error}
\end{figure*}

In Table~\ref{tab:rings-3dbarolo} we provide the best-fit values for each ring for the free parameters: rotational velocity, dispersion velocity, inclination, position angle and system velocity. In Figure~\ref{fig:trial-and-error} we display the residuals (weighted by the moment-0 map) for the different inclination for the kinematics modeling.

\begin{table*}[!h]
\label{tab:rings-3dbarolo}    
\caption{Best-fit parameters for each ring as a  \barolo output.}

\centering          
\begin{tabular}{l c c c c c}     
\hline\hline       

Ring   &   v$_{\rm rot}$ (km s$^{-1}$)  &   v$_{\rm disp}$ (km s$^{-1}$)       & INC (deg)        &   P.A. (deg)          &  V$^{sys}$ (km s$^{-1}$)       \\ 
\hline 
1      &   104$_{-12}^{+10}$            & 15$_{-6}^{+7}$                        & 24$_{-3}^{+3}$  &   179$_{-4}^{+4}$      &  -9$_{-5}^{+5}$   \\ 
2      &   110$_{-21}^{+25}$            & 28$_{-7}^{+10}$                       & 21$_{-8}^{+10}$  &   170$_{-10}^{+9}$     &  -15$_{-4}^{+5}$ \\  
3      &   44$_{-6}^{+5}$               & 36$_{-13}^{+9}$                       & 25$_{-3}^{+5}$  &   182$_{-5}^{+4}$      &  -1$_{-8}^{+8}$  \\ 
\hline  
\hline
\end{tabular}
\end{table*}




\end{document}